\documentclass[%
superscriptaddress,
preprint,
 amsmath,amssymb,
 aps,
]{revtex4-2}

\usepackage{comment}
\usepackage{graphicx, grffile}
\usepackage{dcolumn}
\usepackage{physics}
\usepackage{bm}
\usepackage{hyperref}
\usepackage{cleveref}
\usepackage{fancyhdr}
\usepackage{romannum}
\usepackage{enumerate}
\usepackage{natbib}
\usepackage{tabularx}
\usepackage{floatrow}
\usepackage{subfig}
\usepackage{caption}
\usepackage{setspace}
\usepackage{comment}

\newcommand{\myref}[1]{(\ref{#1})}

\graphicspath{{figure_ms/}}

\begin{document}

\renewcommand{\thepage}{\arabic{page}}

\title{Fluctuations and Persistence in Quantum Diffusion on Regular Lattices}

\author{Cheng Ma}
\affiliation{Department of Physics, Applied Physics and Astronomy, Rensselaer Polytechnic Institute, \\
Troy, NY 12180, USA }
\affiliation{Network Science and Technology Center, Rensselaer Polytechnic Institute, Troy, NY 12180, USA}
\author{Omar Malik}
\affiliation{Department of Physics, Applied Physics and Astronomy, Rensselaer Polytechnic Institute, \\
Troy, NY 12180, USA }
\affiliation{Network Science and Technology Center, Rensselaer Polytechnic Institute, Troy, NY 12180, USA}

\author{G. Korniss}
\affiliation{Department of Physics, Applied Physics and Astronomy, Rensselaer Polytechnic Institute, \\
Troy, NY 12180, USA }
\affiliation{Network Science and Technology Center, Rensselaer Polytechnic Institute, Troy, NY 12180, USA}


\begin{abstract}
We investigate quantum persistence by analyzing amplitude and phase fluctuations of the wave function governed by the time-dependent free-particle Schr\"odinger equation. The quantum system is initialized with local random uncorrelated Gaussian amplitude and phase fluctuations.
In analogy with classical diffusion, the persistence probability is defined as the probability that the local (amplitude or phase) fluctuations have not changed sign up to time $t$. Our results show that the persistence probability in quantum diffusion exhibits {\em exponential-like tails}. More specifically, in $d$$=$$1$ the persistence probability decays in a stretched exponential fashion, while in $d$$=$$2$ and $d$$=$$3$ as an exponential.
We also provide some insights by analyzing the two-point spatial and temporal correlation functions in the limit of small fluctuations. In particular, in the long-time limit, the temporal correlation functions for both local amplitude and phase fluctuations become {\em time-homogeneous}, i.e., the zero-crossing events correspond to those of a {\em stationary Gaussian process}, with sufficiently fast-decaying power-law tail of its autocorrelation function, implying an exponential-like tail of the persistence probabilities.
\end{abstract}

\maketitle

\section{Introduction}

Classical persistence has been widely studied in a variety of diffusive and interacting systems \cite{Bray2013}. In diffusive systems, the persistence probability is defined as the probability that the relevant field (e.g., the diffusive field or the density fluctuations) has stayed above the mean or has not reached a threshold at a location up to time $t$ \cite{Majumdar_PRL1996,newman98,ben-naim2014}. More generally, including interacting systems or fluctuating interfaces, the persistence probability can be defined as the probability that a location or region remained ``active", or a spin has not flipped, or the interface has not crossed zero up to time $t$ \cite{Derrida_JPA1994,Derrida_PRL1995,Howard_JPA1998,Silva2004,Majumdar_PRL1996b,Majumdar_PRL1996c,Fuchs_2008,Grassberger_JSM2009,Toro_PRE1999,Krug_PRE1997}.
Even for stationary Gaussian processes, calculating the persistence probability  (or the first-passage time distribution or zero-crossing probabilities) of the stochastic variable is a non-trivial problem \cite{Rice_1944,Rice_1945,Newell_1962,Slepian_1962,Derrida_PRL1996}. Understanding the temporal and finite-size scaling properties of the persistence probability has a wide range of applications from classical materials transport to epidemic spreading and opinion dynamics in social networks. It was only recently explored what the fundamental effects of disorder and random/complex network structures are on classical diffusive persistence \cite{Malik_PRE2024}.

The time scales associated with the spatial spreading of initially localized density fluctuations are well understood in classical and quantum diffusion. In particular, the width of a Gaussian density profile in classical diffusion grows as $\sqrt{\langle x^2 \rangle}\sim t^{1/2}$, while the width of a Gaussian wave packet in quantum diffusion scales as  $\sqrt{\langle x^2 \rangle}\sim t$ (for long times) \cite{Townsend_2012} and both maintaining their initial Gaussian shape.

It is also of interest to study the relaxation properties of local density fluctuations in classical and quantum diffusion with {\em spatially-distributed random} initial conditions. For classical diffusion, the persistence probability is defined as the probability that the diffusive field or density fluctuations at some site $\bm{x}$ has stayed above (or below) zero between time $0$ and time $t$. For translationally invariant systems with spatially-random initial conditions, this probability is independent of the site and has been shown to scale as a power law, $P(t)\sim t^{-\theta}$ \cite{Bray2013}, where $\theta$ is the persistence exponent and depends on the topology and the dimension of the underlying systems.
Recalling the properties of the two-point correlation functions in classical diffusion provides some insight into the origin of this power-law decay.
Being a linear partial differential equation for the local field variable $\phi(\bm{x},t)$, $\partial_t \phi = a \nabla^2\phi$ ($a>0$), the classical diffusion equation can be easily solved for arbitrary initial conditions, including uncorrelated random initial disturbances with zero mean for all sites and $\langle \phi(\bm{x},0)\phi(\bm{x}',0)\rangle =D\delta(\bm{x}-\bm{x}')$ ($D>0$, and the initial distribution can be Gaussian for concreteness). Further, one can also obtain the two-point correlation functions, e.g., $\langle \phi(\bm{x},t)\phi(\bm{x}',t)\rangle =D(8\pi a t)^{-d/2}\exp{-\frac{(\bm{x}-\bm{x}')^2}{8at}}$ and $\langle \phi(\bm{x},t)\phi(\bm{x},t')\rangle =D(4\pi a (t+t'))^{-d/2}$ \cite{Bray2013,Majumdar_PRL1996}.
The latter (temporal) correlations show that the original Gaussian process associated with classical diffusion and random uncorrelated initial condition is {\em not} a stationary (or time-homogeneous) process. However, the normalized temporal autocorrelation function in {\em logarithmic} time is a stationary process (i.e., only depends on the difference of the new ``time" variables) with an exponential tail \cite{Majumdar_PRL1996,Bray2013}. From Newell and Rosenblatt's 1962 result \cite{Newell_1962} it then follows \cite{Bray2013} that the persistence probability decays exponentially in the logarithmic time variable, hence as a power law in the original one. However, the value of the power-law exponent (the rate of the exponential decay in logarithmic time) is non-trivial, and depends on the full stationary autocorrelator. With the exception of two-dimension, $\theta_2=3/16=0.1875$ \cite{Schehr2018,Dornic2018}, only approximate results are available , e.g., $\theta_1\simeq 0.1207$ and $\theta_3\simeq 0.2380$, in one and three dimensions, respectively \cite{Majumdar_PRL1996,Derrida_PRL1996,newman98}.
On two-dimensional disordered lattices with bond disorder, as shown by simulations, the persistence exponent changes to $\theta\simeq 0.141$ at around the percolation threshold \cite{Malik_PRE2024}. On other types of random graphs, e.g., on ER networks, scaling is less clear, if at all exists \cite{Malik_PRE2024}.

In this paper, we study local fluctuations and persistence in quantum diffusion and ask questions closest in analogy to classical diffusion. Applicable to quantum dynamics, we consider local fluctuations for the amplitude and phase of the wave function, and analyze their persistence probabilities, i.e., the probability that these fluctuations have not changed sign up to time $t$. Note that for small fluctuations, the fluctuations of the amplitude of the wave function are proportional to those of the (probability) density. We will also provide some insights to the coupled amplitude and phase fluctuations by investigating their relevant two-point correlation functions. Note that Goswami and Sen studied persistence probability in a discrete quantum random walk \cite{Goswami_PRE2010}, i.e., the probability that a quantum random walker starting at the origin has not reached a given site by time $t$, a different aspect of quantum diffusion. Our study addresses the local spatio-temporal characteristics of the free-particle wave function initiated from random translationally invariant and uncorrelated initial amplitude and phase distributions.

\section{Numerical framework}
\label{sec:numerical}

For a single non-relativistic particle in the absence of potential energy, the time evolution of the wavefunction is described by the Schr\"{o}dinger's equation, where the Hamiltonian $\mathcal{H} = -\frac{\hbar^2}{2m}\nabla^2 $, and $m$ is the mass of the particle. Given an initial state $\psi(\bm{x}, 0)$, the wave function $\psi(\bm{x},t)$ evolves deterministically according to
\begin{equation}
		i\hbar \partial_t \psi  = \mathcal{H} \psi.
\label{eq:se_pde}
\end{equation}
In order to numerically integrate Eq.~\myref{eq:se_pde}, we have implemented the Crank-Nicholson scheme \cite{Iitaka_PRE1994,Robertson_2011,Khan_2022},
	\begin{equation}
		\psi(\bm{x},t+\Delta t) = \left(1 + \frac{i \Delta t \mathcal{H} }{ 2 	\hbar}\right)^{-1} \left(1 - \frac{i \Delta t \mathcal{H} }{ 2 	 \hbar}\right) \psi(\bm{x},t)  \;,
  \label{eq:se_discretized}
	\end{equation}
which is a unitary and unconditionally stable time-discretization scheme. In this paper, we will refer to this method as the exact numerical integration (ENI).
For the Laplacian, we employed standard spatial discretization, e.g., in one dimension, $\nabla^2\psi(x,t)\simeq[\psi(x_{i+1},t)+\psi(x_{i-1},t)-2\psi(x_{i},t)]/(\Delta x)^2$. To set convenient scales for numerical explorations, we have $m=m_e=9.1\times 10^{-31}{\rm kg}$. Then $\alpha\equiv\frac{\hbar}{2m_e}=5.794\times 10^{-2} ({\rm nm})^2{\rm fs}^{-1}$. In this paper, the unit of length is $1 {\rm nm}$ and the unit of time is $1 {\rm fs}$ everywhere. In these units, the time discretization is $\Delta t=$$1$, and unless stated otherwise in figure captions, $\Delta x$$=$$1$. We employ $d$-dimensional regular lattices with  $N=(L/\Delta x)^d = 10000$ sites and periodic boundary conditions. The number of realizations of the random initial conditions $M =10$ (see below).

Note that here we implemented the above Crank-Nicholson time discretization scheme \cite{Iitaka_PRE1994,Robertson_2011,Khan_2022} [Eq.~\myref{eq:se_discretized}] because it can be readily employed for more complicated systems and future studies, including the presence of potentials, non-linear Schr\"{o}dinger equations, and disordered lattices, where the exact continuous time-evolution (ECTE) solution is not available. Further, since in our case, the continuous-time evolution for the free particle Schr\"{o}dinger equation is well known, we were able to perform the numerical precision test and comparison between the two methods (ENI and ECTE) for the random initial conditions described below, and illustrated in Supplemental Material/I. For example, on a one-dimensional regular lattice with periodic boundary conditions, the ECTE solution is
\begin{equation}
\psi(x,t) = \frac{1}{L}  \sum_{k} \tilde{\psi}_{k}(0)
e^{-\frac{i}{\hbar}E_{k}t} e^{i k x} \;,
\label{eq:ECTE_solution}
\end{equation}
with $k=\frac{2\pi n}{N\Delta x}$,
$n=-\frac{N}{2}+1, -\frac{N}{2}+2, \ldots, \frac{N}{2}$ (using even $N$),
and energy eigenvalues
\begin{equation}
E_{k} = \frac{\hbar^2}{m (\Delta x)^2} ( 1-\cos(k\Delta x) ) \;,
\label{eq:ECTE_eigenvalue}
\end{equation}
and $\tilde{\psi}_{k}(0)$ are the Fourier components of the initial wave function.

\subsection*{Initial wave function with random uncorrelated amplitude and phase fluctuations}

The uniform-amplitude wave function with arbitrary uniform phase, $\psi_o=R_o e^{i\vartheta_o}$ where $R_o=L^{-d/2}$ is a solution of the time-independent Schr\"{o}dinger equation; it physically corresponds to the ground state of a ``free particle" in a box with periodic boundary conditions. It is also a trivial solution of the time-dependent Schr\"{o}dinger equation \myref{eq:se_pde}. In this paper, we focus on small fluctuations about this ``fixed-point" solution (and chose $\vartheta_o =0$ without loss of generality). Our variables of our interest are the amplitude $R(\bm{x},t)$ and the phase $\vartheta(\bm{x},t)$ of the complex wave function, $\psi(\bm{x},t)=R(\bm{x},t)e^{i\vartheta(\bm{x},t)}$. Further, we describe the amplitude of the wave function in terms of the fluctuations $r(\bm{x},t)$ about the uniform state $R_o$,
\begin{equation}
    R(\bm{x},t)=R_o +  r(\bm{x},t) \;.
\end{equation}
We also introduce the dimensionless variable  $u(\bm{x},t) \equiv r(\bm{x},t)/R_o$, the relative size of the amplitude fluctuations. We characterize the initial state in terms of the variance of the local amplitude and phase fluctuations. More specifically, for the initial conditions on the discretized lattice, we employed uncorrelated Gaussian random variables for the initial amplitude and phase fluctuations, $\langle u(\bm{x},0)\rangle=0$, $\langle \vartheta(\bm{x},0)\rangle=0$, $\langle u(\bm{x},0)u(\bm{x}',0)\rangle=\sigma_u^2 \delta_{\bm{x},\bm{x}'}$, $\langle \vartheta(\bm{x},0)\vartheta(\bm{x}',0)\rangle=\sigma_{\vartheta}^2\delta_{\bm{x},\bm{x}'}$, and  $\langle u(\bm{x},0)\vartheta(\bm{x}',0)\rangle=0$. In our subsequent notations, we will be using
$\sigma_u^2(t)\equiv\langle u^2(\bm{x},t)\rangle$, $\sigma_{\vartheta}^2(t)\equiv\langle\vartheta^2(\bm{x},t)\rangle$ for the local time-dependent variances, and reserve the notations
$\sigma_u^2\equiv\sigma_u^2(0)=\langle u^2(\bm{x},0)\rangle$, $\sigma_{\vartheta}^2\equiv\sigma_{\vartheta}^2(0)=\langle\vartheta^2(\bm{x},0)\rangle$ for the local initial variances. Note that after averaging over the initial distribution of the random amplitude and phase variables $\langle\ldots\rangle$, the system is translationally invariant, hence there will be no site dependence of these fluctuations at any time.

Before we present and discuss the results obtained by the exact numerical integration (ENI) of Eq.~\myref{eq:se_discretized} (Sec.~\ref{sec:numerical_results}), we provide some insight by studying the limit of small-amplitude fluctuations of this system in the next Section.

\section{Small-amplitude approximation for the time evolution and the two-point correlation functions}
\label{sec:analytic}

While the time-dependent Schr\"odinger equation is a linear partial differential equation for the complex wave function $\psi(\bm{x},t)=R(\bm{x},t)e^{i\vartheta(\bm{x},t)}$ [Eq.~\myref{eq:se_pde}], it is a non-linear one for the variables of our interest $R(\bm{x},t)$ and $\vartheta(\bm{x},t)$. Starting from Eq.~\myref{eq:se_pde} (see, e.g., \cite{Sakurai_2021}),
\begin{equation}
    i\hbar \partial_t
    \left[ R(\bm{x},t)e^{i\vartheta(\bm{x},t)}\right]
    = -\frac{\hbar^2}{2m}\nabla^2 \left[ R(\bm{x},t)e^{i\vartheta(\bm{x},t)} \right]   \;,
\end{equation}
one has
\begin{equation}
    i\hbar \left[\partial_t R + i R\partial_t\vartheta \right] =
    -\frac{\hbar^2}{2m}\left[ \nabla^2 R + 2i(\nabla R)(\nabla\vartheta)
    - R(\nabla\vartheta)^2 + i R\nabla^2\vartheta
    \right]
    \label{eq:sch_R_theta}
    \;\;\;,
\end{equation}
or in terms of the fluctuations of the amplitude $r(\bm{x},t)$, $R(\bm{x},t)=R_o +  r(\bm{x},t)$,
\begin{equation}
    i\hbar \left[\partial_t r + i(R_o+r)\partial_t\vartheta \right] =
    -\frac{\hbar^2}{2m}\left[ \nabla^2 r + 2i(\nabla r)(\nabla\vartheta)
    -(R_o+r)(\nabla\vartheta)^2 + i(R_o+r)\nabla^2\vartheta
    \right]
    \label{eq:sch_r_theta}
    \;.
\end{equation}
The above forms of Schr\"odinger's equation are standard starting points to study quantum turbulence \cite{Nore_1997,Chiueh_2011} (with fluid density $\rho$$=$$R^2$ and velocity $\bm{v}\propto\nabla\vartheta$), while in the Gross-Pitaevskii equation (a nonlinear Schr\"odinger equation) it is commonly employed to study the macroscopic wave function and its fluctuations as excitations in Bose-Einstein condensates \cite{Salazar_EJP2013,Hellweg_APB2001}. Our reason here to employ this framework is simply that the above equation directly involves the variables of our interest: the amplitude and the phase fluctuations of the wave function. Here we focus on small fluctuations about the uniform amplitude and uniform phase, such that $|u(\bm{x},t)| = |r(\bm{x},t)/R_o|\ll 1$, and  $|\vartheta(\bm{x},t)| \ll 1$. Note that in this limit $|\psi(\bm{x},t)|^2=(R_o +  r(\bm{x},t))^2\simeq R_o^2(1+2r(\bm{x},t)/R_o)=R_o^2(1+2u(\bm{x},t))$, hence, the amplitude fluctuations are proportional to the (probability) density fluctuations. Linearizing Eq.~\myref{eq:sch_r_theta}, we obtain
\begin{equation}
	\begin{split}
        \partial_t u & \simeq - \alpha \nabla ^2 \vartheta \\
		\partial_t \vartheta & \simeq \alpha \nabla ^2 u
    \label{eq:u_coupled}
	\end{split}
\end{equation}
for the imaginary and real parts, respectively, where $\alpha\equiv\frac{\hbar}{2m}>0$. The above set of coupled linear equations for the amplitude and phase fluctuations can be solved for arbitrary initial conditions in momentum space, i.e., using spatial Fourier transforms (see Appendix~\ref{appendix_A}).

\subsection{Equal-time spatial correlation functions}

Here we employ uncorrelated Gaussian random variables for the initial conditions, $u(\bm{x},0)=\eta(\bm{x})$ and $\vartheta(\bm{x},0)=\xi(\bm{x})$, such that
$\langle \eta(\bm{x})\rangle=0$, $\langle \xi(\bm{x})\rangle=0$, $\langle \eta(\bm{x})\eta(\bm{x}')\rangle=D_u \delta(\bm{x}-\bm{x}')$, $\langle \xi(\bm{x})\xi(\bm{x}')\rangle=D_{\vartheta}\delta(\bm{x}-\bm{x}')$, and  $\langle \eta(\bm{x})\xi(\bm{x}')\rangle=0$. The interpretation of the $d$-dimensional delta-function in connection with the spatial discretization scheme is $\delta(\bm{x}-\bm{x}')\simeq\delta_{\bm{x},\bm{x}'}/(\Delta x)^d$, hence $\sigma_u^2=D_u/(\Delta x)^d $ and $\sigma_{\vartheta}^2=D_{\vartheta}/(\Delta x)^d$.
One can then obtain the relevant two-point correlation functions. As shown in Appendix~\ref{appendix_B}, the equal-time spatial correlations can be written as
\begin{equation}
\begin{split}
& \langle u(\bm{x},t) u(\bm{x}',t)  \rangle  \simeq
\frac{D_{u} + D_{\vartheta}}{2} \frac{\delta_{\bm{x},\bm{x}'}}{(\Delta x)^d}
+  \frac{D_{u} - D_{\vartheta}}{2} \frac{1}{(8\pi\alpha t)^{d/2}}
   \Re \left( \prod_{j=1}^{d} F(x_j - x_j',t;\Delta x) \right)  \\
& \langle \vartheta(\bm{x},t) \vartheta(\bm{x}',t)  \rangle  \simeq
\frac{D_{u} + D_{\vartheta}}{2} \frac{\delta_{\bm{x},\bm{x}'}}{(\Delta x)^d}
- \frac{D_{u} - D_{\vartheta}}{2} \frac{1}{(8\pi\alpha t)^{d/2}}
   \Re \left( \prod_{j=1}^{d} F(x_j - x_j',t;\Delta x) \right)  \\
& \langle u(\bm{x},t) \vartheta(\bm{x}',t)  \rangle  \simeq
 \frac{D_{u} - D_{\vartheta}}{2} \frac{1}{(8\pi\alpha t)^{d/2}}
   \Im \left( \prod_{j=1}^{d} F(x_j - x_j',t;\Delta x) \right)
\end{split} \;\;\;,
\label{corr_spatial_discrete}
\end{equation}
where
\begin{equation}
F(y,t;\Delta x) \equiv
\frac{ e^{i\frac{y^2}{8\alpha t} -i\frac{\pi}{4} }  }{2}
\left[
\erf\left(\sqrt{2i\alpha t}(\frac{\pi}{\Delta x}+ \frac{y}{4\alpha t}) \right)
-
\erf\left(\sqrt{2i\alpha t}(-\frac{\pi}{\Delta x}+ \frac{y}{4\alpha t}) \right) \right]  \;\;\;.
\end{equation}
We refer to the above results as the linearized large system-size limit (LLSSL). In the limit of $\Delta x\to 0$, the above expressions become (Appendix~\ref{appendix_B})
		\begin{equation}
		\begin{split}
			\langle u(\bm{x},t) u(\bm{x}',t)  \rangle
			& \simeq  \frac{D_{u} + D_{\vartheta}}{2} \delta(\bm{x}-\bm{x}') +
   \frac{D_{u} - D_{\vartheta}}{2} \frac{1}{(8 \pi\alpha t)^{d/2}} \cos( \frac{(\bm{x}-\bm{x}')^2}{8 \alpha t} - \frac{d\pi}{4}) \\
\langle \vartheta(\bm{x},t) \vartheta(\bm{x}',t)  \rangle
			& \simeq   \frac{D_{u} + D_{\vartheta}}{2} \delta(\bm{x}-\bm{x}') -
   \frac{D_{u} - D_{\vartheta}}{2} \frac{1}{(8 \pi\alpha t)^{d/2}} \cos( \frac{(\bm{x}-\bm{x}')^2}{8 \alpha t} - \frac{d\pi}{4}) \\
\langle u(\bm{x},t) \vartheta(\bm{x}',t)  \rangle
			& \simeq   \frac{D_{u} - D_{\vartheta}}{2} \frac{1}{(8 \pi\alpha t)^{d/2}} \sin( \frac{(\bm{x}-\bm{x}')^2}{8 \alpha t} - \frac{d\pi}{4})
		\end{split} \;\;\;.
        \label{eq:spatial_corr}
		\end{equation}
We refer to this limit as the linearized large system-size continuum limit (LLSSCL). Note that while Eq.~\myref{eq:spatial_corr} neglects both spatial discretization and non-linearities from Eq.~\myref{eq:sch_r_theta} to Eq.~\myref{eq:u_coupled}, it provides the basic trend of the spatial correlation functions, and will be compared to the results obtained by exact numerical integration [Eq.~\myref{eq:se_discretized}].

A number of interesting properties immediately follow from the above expressions, Eqs.~\myref{corr_spatial_discrete} and  \myref{eq:spatial_corr}(regardless of discretization). First,
\begin{equation}
\langle u(\bm{x},t) u(\bm{x}',t) \rangle + \langle \vartheta(\bm{x},t) \vartheta(\bm{x}',t) \rangle
=
(D_{u} + D_{\vartheta}) \frac{\delta_{\bm{x},\bm{x}'}}{(\Delta x)^d}
=
(\sigma^2_u + \sigma^2_{\vartheta}) \delta_{\bm{x},\bm{x}'} \;\;\; .
\label{sigma_sum}
\end{equation}
i.e., independent of time. In particular, for $\bm{x}=\bm{x}'$, Eq.~\myref{sigma_sum} becomes
\begin{equation}
\sigma_u^2(t) + \sigma_{\vartheta}^2(t)	
=
\langle u^2(\bm{x},t) \rangle + \langle \vartheta^2(\bm{x},t) \rangle
=
\sigma^2_u + \sigma^2_{\vartheta} = \mathrm{const.}
\;\;\;,
\label{eq:sum_of_variances}
\end{equation}
or equivalently,
\begin{equation}
\frac{\sigma_u^2(t) + \sigma_{\vartheta}^2(t)}
{\sigma^2_u + \sigma^2_{\vartheta}} = 1
\;\;\;.
\label{eq:sum_of_variances_scaled}
\end{equation}
Further, from Eqs.~\myref{corr_spatial_discrete} and \myref{eq:spatial_corr} it also follows that
	 \begin{equation}
  \begin{split}
  \sigma_u^2(\infty) 	 & =
  \frac{D_{u} + D_{\vartheta}}{2} \frac{1}{(\Delta x)^d} =
  \frac{\sigma^2_u + \sigma^2_{\vartheta}}{2}  \\
\sigma_{\vartheta}^2(\infty)  	 & =
  \frac{D_{u} + D_{\vartheta}}{2} \frac{1}{(\Delta x)^d} =
  \frac{\sigma^2_u + \sigma^2_{\vartheta}}{2}
  \end{split} \;\;\;.
  \label{eq:asymptotic_variance}
	\end{equation}
The above expressions show that average local amplitude (and phase) fluctuations do not asymptotically vanish in the long-time limit. Unlike classical diffusion where density fluctuation for each wave number is governed by exponential relaxation in time with decay rate $a{\bm k}^2$, in quantum diffusion the time evolution of each wave-number is oscillatory with a quadratic dispersion $\omega_{\bm k}=\alpha {\bm k}^2$ (as becomes also explicit at the level of small-amplitude oscillations, Appendix~\ref{appendix_A}). The superposition of these propagating waves with random spatially-distributed initial conditions ultimately gives rise to the above asymptotic non-zero average local amplitude and phase fluctuations.

One intriguing feature of the above coupled amplitude-phase behavior is the special case when the variance of the initial random variables are identical, $D_u$$=$$D_{\vartheta}$ ($\sigma^2_u$$=$$\sigma^2_{\vartheta}$): the quantum system becomes spatially correlation-free at all times, as suggested by Eq.~\myref{corr_spatial_discrete} and  \myref{eq:spatial_corr}.

\subsection{Single-site temporal correlation functions (autocorrelations)}

For the single-site temporal correlation functions we find (Appendix~\ref{appendix_C})
\begin{equation}
\begin{split}
& \langle u(\bm{x},t) u(\bm{x},t')  \rangle  \simeq \\
& \frac{D_{u} + D_{\vartheta}}{2} \frac{1}{(4\pi\alpha |t-t'|)^{d/2}}   \Re \left( H^d(t-t';\Delta x) \right)
+
\frac{D_{u} - D_{\vartheta}}{2} \frac{1}{(4\pi\alpha (t+t'))^{d/2}}   \Re \left( H^d(t+t';\Delta x) \right)  \\
& \langle \vartheta(\bm{x},t) \vartheta(\bm{x},t')  \rangle  \simeq \\
& \frac{D_{u} + D_{\vartheta}}{2} \frac{1}{(4\pi\alpha |t-t'|)^{d/2}}   \Re \left( H^d(t-t';\Delta x) \right)
-
\frac{D_{u} - D_{\vartheta}}{2} \frac{1}{(4\pi\alpha (t+t'))^{d/2}}   \Re \left( H^d(t+t';\Delta x) \right)
\\
& \langle u(\bm{x},t) \vartheta(\bm{x},t')  \rangle  \simeq \\
& -\frac{D_{u} + D_{\vartheta}}{2} \frac{\mathrm{sgn}^{d}(t-t')}{(4\pi\alpha |t-t'|)^{d/2}}   \Im \left( H^d(t-t';\Delta x) \right)
+
\frac{D_{u} - D_{\vartheta}}{2} \frac{1}{(4\pi\alpha (t+t'))^{d/2}}   \Im \left( H^d(t+t';\Delta x) \right)
\end{split} \;\;\;,
\label{corr_temporal_discrete}
\end{equation}
where
\begin{equation}
H(\tau;\Delta x) \equiv
e^{-i\frac{\pi}{4} }
\erf\left(\sqrt{i\alpha\tau}(\frac{\pi}{\Delta x}) \right)  \;\;\;.
\end{equation}
In the limit of $\Delta x\to 0$, the above expressions approach (Appendix~\ref{appendix_C})
		\begin{equation}
		\begin{split}
\langle u(\bm{x}, t) u(\bm{x}, t')  \rangle
			& \simeq \frac{D_{u} + D_{\vartheta}}{2}  \frac{\cos(\frac{d\pi}{4})}{(4\pi\alpha |t-t'|)^{d/2}}
        + \frac{D_{u} - D_{\vartheta}}{2}  \frac{\cos(\frac{d\pi}{4})}{(4\pi\alpha (t+t'))^{d/2}}  \\
\langle \vartheta(\bm{x}, t) \vartheta(\bm{x}, t')  \rangle
			& \simeq  \frac{D_{u} + D_{\vartheta}}{2}  \frac{\cos(\frac{d\pi}{4})}{(4\pi\alpha |t-t'|)^{d/2}}
        - \frac{D_{u} - D_{\vartheta}}{2}  \frac{\cos(\frac{d\pi}{4})}{(4\pi\alpha (t+t'))^{d/2}}  \\
\langle u(\bm{x}, t) \vartheta(\bm{x}, t')  \rangle
			& \simeq   \frac{D_{u} + D_{\vartheta}}{2}  \frac{\sin(\frac{d\pi}{4})}{(4\pi\alpha |t-t'|)^{d/2}} \mathrm{sgn}^{d}(t-t')
        - \frac{D_{u} - D_{\vartheta}}{2}  \frac{\sin(\frac{d\pi}{4})}{(4\pi\alpha (t+t'))^{d/2}}
		\end{split} \;\;\;,
        \label{eq:temporal_corr}
		\end{equation}
where $\mathrm{sgn}(\tau)$ is the sign function. As can be seen from the above expressions (regardless of discretization), in the long-time limit, $t=t'+\tau$ with $\tau$ fixed and $t'\to \infty$, the second terms in the above expressions vanish, hence these correlation functions become time-homogeneous, i.e., will only depend on $\tau$, $C_{u}(\tau,t')\equiv\langle u(\bm{x},t'+\tau) u(\bm{x},t')\rangle\simeq \hat{C}_{u}(\tau)$ and
$C_{\vartheta}(\tau,t')\equiv\langle \vartheta(\bm{x},t'+\tau) \vartheta(\bm{x},t')\rangle\simeq \hat{C}_{\vartheta}(\tau)$.
Specifically, for finite discretization,
\begin{equation}
\hat{C}_{u}(\tau) \simeq \hat{C}_{\vartheta}(\tau)
\simeq
\frac{D_{u} + D_{\vartheta}}{2} \frac{1}{(4\pi\alpha |\tau|)^{d/2}}   \Re \left( H^d(\tau;\Delta x) \right)
\;\;\;.
\label{autocorr_asymp}
\end{equation}
Further, for the special case $D_u$$=$$D_{\vartheta}$ ($\sigma^2_u$$=$$\sigma^2_{\vartheta}$), these temporal correlations are time-homogeneous for all times, as can be seen from Eq.~\myref{corr_temporal_discrete}.

\subsection{Amplitude and Phase Persistence}

The important connections to the persistence probabilities of the amplitude and phase $P_u(t)$ and $P_{\vartheta}(t)$ (the probabilities that the local amplitude and phase fluctuations have not changed sign up to time $t$) are given by the form of the temporal correlation functions Eqs.~\myref{corr_temporal_discrete} and \myref{eq:temporal_corr}: within the framework of linearized fluctuations, in the long-time limit they are {\em time-homogeneous} (i.e., depend only on $\tau=t-t'$) and decay as $\sim|\tau|^{-d/2}$. (The long-time behavior of the autocorrelation functions [Eq.~\myref{corr_temporal_discrete}] can be obtained utilizing the asymptotic expansion of the error function Eq.~\myref{autocorr_asymptotic}.)
Thus, the persistence probability of small amplitude and phase fluctuations in quantum diffusion is equivalent to that of a stationary Gaussian process with this type of tail of the autocorrelation function. Newell and Rosenblatt \cite{Newell_1962} showed that the persistence probability for precisely these types of autocorrelations decays {\em faster than any power law}.
Further, based on the upper and lower bounds specific to the value of the exponent of the decay of the autocorrelation function \cite{Newell_1962}, and some recent results \cite{Feldheim_2014,Dembo_2016,Feldheim_arxiv2021}, we anticipate that in $d$$=$$1$, the persistence probabilities will decay as a stretched exponential and for $d$$=$$2$ and $d$$=$$3$ as an exponential.

Given the form of Schr\"odinger's equation \myref{eq:se_discretized} and subsequently Eqs.~\myref{corr_temporal_discrete}, it is clear that for a given initial distribution with $\sigma_u$ and $\sigma_{\vartheta}$, the persistence probability will be a function of the dimensionless variable $\alpha t/(\Delta x)^2$. Combining the above, we anticipate
\begin{equation}
P_{u,\vartheta}(t) \simeq \exp{ -c\left(\frac{\alpha t}{(\Delta x)^2}\right)^{\beta} }
\;\;\;,
\label{eq:persistence_stretched_exp}
\end{equation}
with $\beta\leq 1$.

\section{Numerical Results and Discussion}
\label{sec:numerical_results}

\subsection{Variance of the local fluctuations}
\begin{figure}[H]
	\centering
		\includegraphics[width=0.9\textwidth]{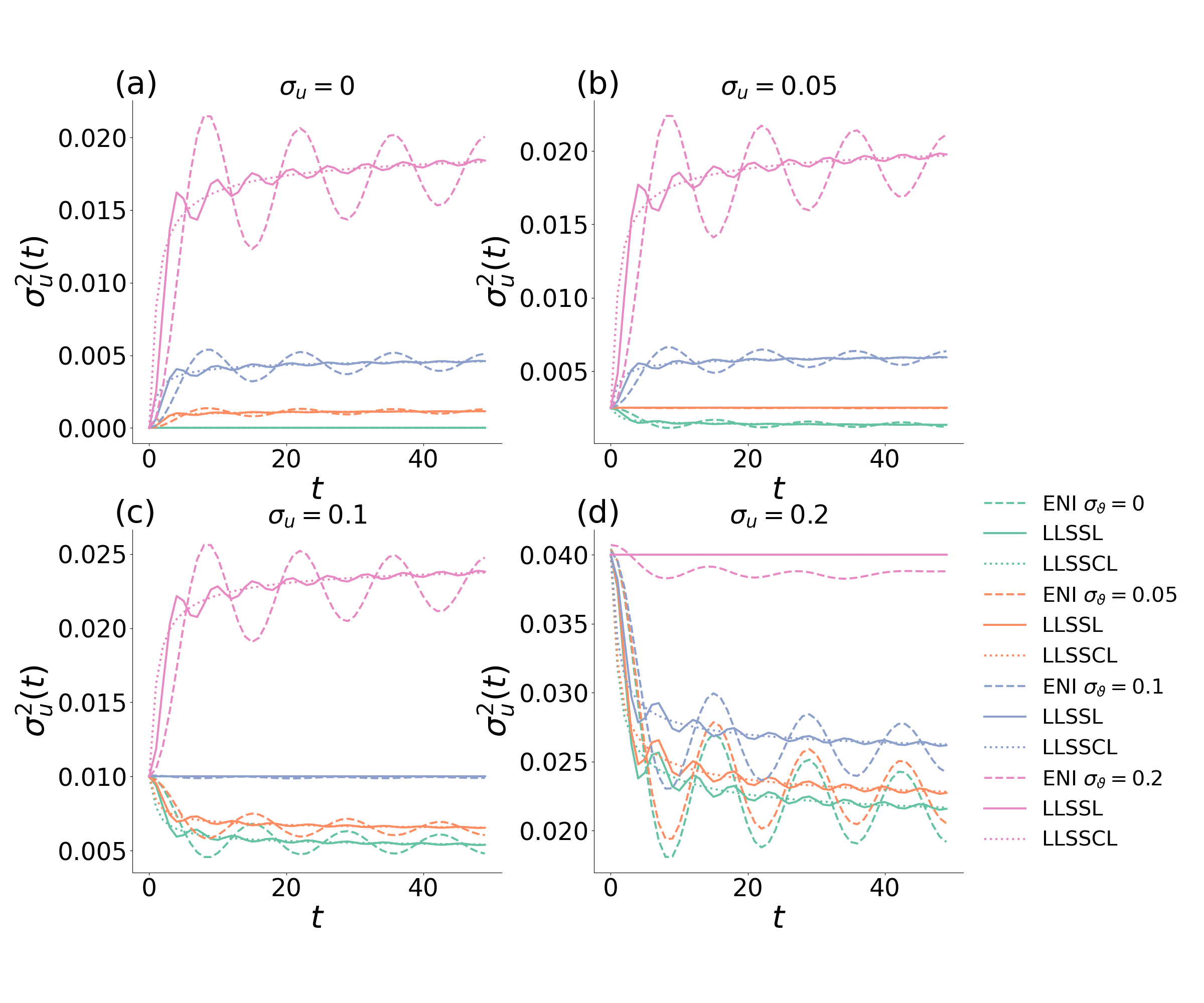}
		\caption{\textbf{Variance of the amplitude fluctuations as a function of time.} In each subfigure, the initial standard deviation $\sigma_u$ is the same. Dashed lines are the exact numerical integration (ENI) of Schr\"{o}dinger equation, solid lines are the analytical prediction by taking the linearized large system-size limit (LLSSL), and dotted lines are the approximation by taking the linearized large system-size and continuum limit (LLSSCL). }
    \label{fig:amplitude_variance}
	\end{figure}

As predicted by the linear approximation [Eq.~\myref{corr_spatial_discrete}], the local variance of both amplitude and phase fluctuations asymptotically approaches a constant value [Eq.~\myref{eq:asymptotic_variance}].
For example, for the local amplitude variance we have
\begin{equation}
\sigma_u^2(t) = \langle u^2(\bm{x},t)\rangle \simeq
\frac{\sigma_{u} + \sigma_{\vartheta}}{2}
+  \frac{\sigma_{u} - \sigma_{\vartheta}}{2} \frac{1}{(8\pi\alpha t/(\Delta x )^2)^{d/2}}    \Re \left( F^d(0,t;\Delta x) \right)
\;\;\;,
\end{equation}
hence
$\sigma_u^2(\infty) = \frac{\sigma_{u} + \sigma_{\vartheta}}{2}$.
We show the numerical results for the amplitude variances in Fig.~\ref{fig:amplitude_variance}.
Our results indicate, in agreement with Eq.~\myref{corr_spatial_discrete}, that in the regime of small fluctuations, for the special case $D_u$$=$$D_{\vartheta}$ ($\sigma^2_u$$=$$\sigma^2_{\vartheta}$), the system becomes ``correlation free", and the single-site variance is independent of time.

\begin{figure}[H]
	\centering
		\includegraphics[width=0.9\textwidth]{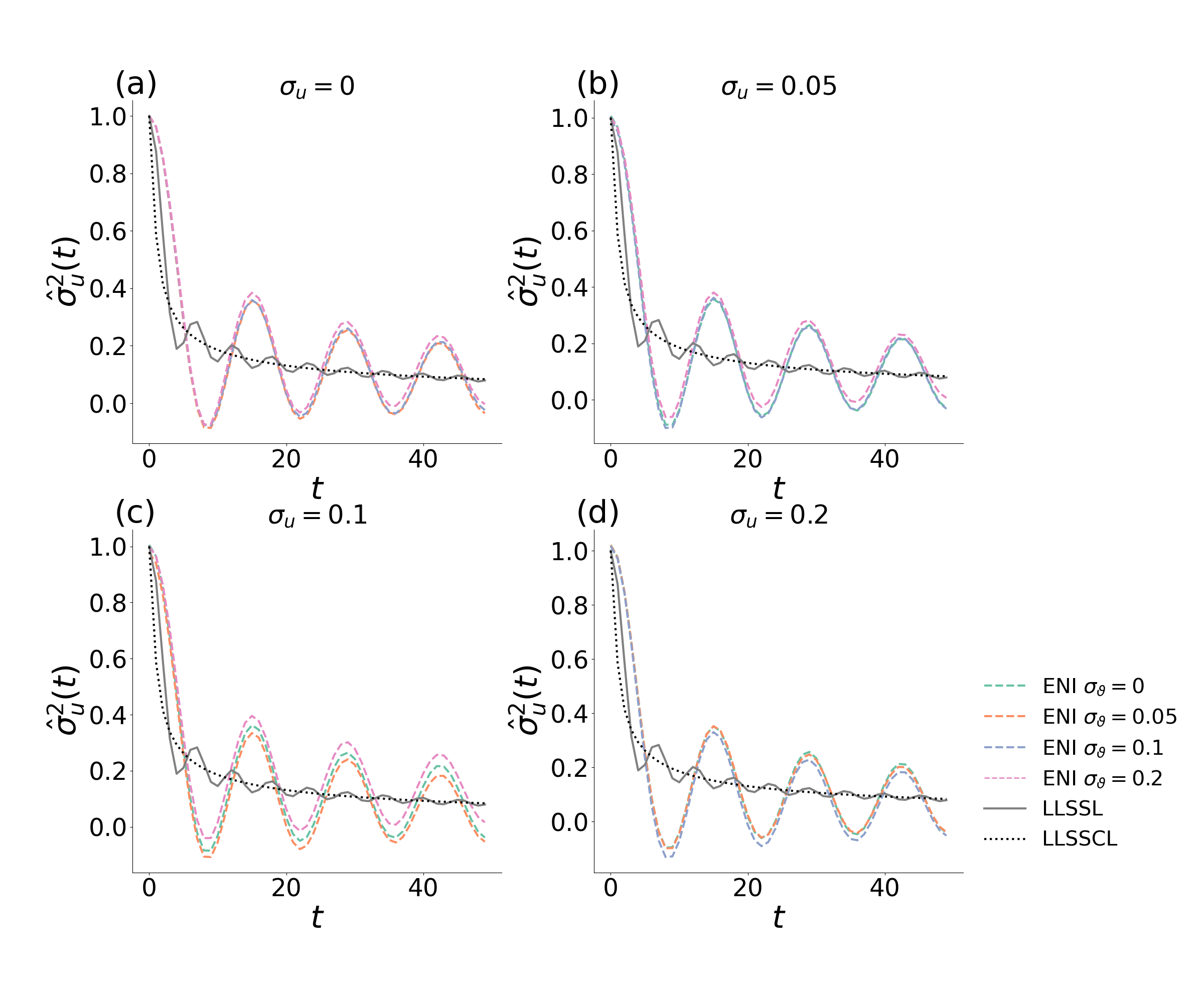}
		\caption{\textbf{Scaled variance of the amplitude fluctuations as a function of time.} In each subfigure, the initial standard deviation $\sigma_u$ is the same. Dashed lines are the exact numerical integration (ENI) of Schr\"{o}dinger equation, solid lines are the analytical prediction by taking the linearized large system-size limit (LLSSL), and dotted lines are the approximation by taking the linearized large system-size and continuum limit (LLSSCL). }
    \label{fig:scaled_amplitude_variance}
	\end{figure}

The linearized approximation [Eq.~\myref{corr_spatial_discrete}] also correctly suggests that the scaled variance (e.g., for the amplitude fluctuations),
\begin{equation}
    \hat{\sigma}_u^2(t) \equiv \frac{\sigma_u^2(t) - \frac{D_u + D_{\vartheta}}{ 2(\Delta x)^d} }{\frac{D_u - D_{\vartheta}}{2(\Delta x)^d} }
    = \frac{\sigma_u^2(t) - \frac{\sigma^2_u + \sigma^2_{\vartheta}}{2} }{\frac{\sigma^2_u - \sigma^2_{\vartheta}}{2} }
=   \frac{1}{\left(8\pi\alpha t/(\Delta x)^2\right)^{d/2}}
   \Re \left( F^d(0,t;\Delta x) \right)
    \label{scaled_variance}
\end{equation}
will collapse on a single curve for various initial values of the variances, as shown in Fig.~\ref{fig:scaled_amplitude_variance}.
The linear approximation, however misses some important details: employing the asymptotic expansion of the error function in Eq.~\myref{scaled_variance} yields an approach to the asymptotic steady-state value as $c_1 t^{-1/2} + c_2 t^{-1}\sin(2\omega_{\max} t)$  [Eq.~\myref{eq:amplitude_variance_asymp}], corresponding to oscillatory corrections with a period of $\pi/\omega_{\max}=\frac{(\Delta x)^2}{\alpha \pi} \approx 5.5$. (Note that $(\pi/\Delta x)^2$ is the largest wave number for a given discretization, hence the corresponding frequency is $\omega_{\max}=\alpha k_{\max}^2=\frac{\alpha\pi^2}{(\Delta x)^2}$.) In contrast, the results of the exact numerical integration (ENI) indicate that the approach to the steady-state exhibits larger amplitude oscillatory decay about the $t^{-1/2}$ trend with a period of about 14. Note that the terms we neglect in Eq.~\myref{eq:sch_r_theta} under linear approximation are heavily dominated by the mode with the largest wave number. Those terms are responsible for the selection of the actual dominant frequency governing the approach to long-time steady-state values and the temporal scaling of the decay of the oscillations.

\begin{figure}[H]
	\centering
		\includegraphics[width=0.9\textwidth]{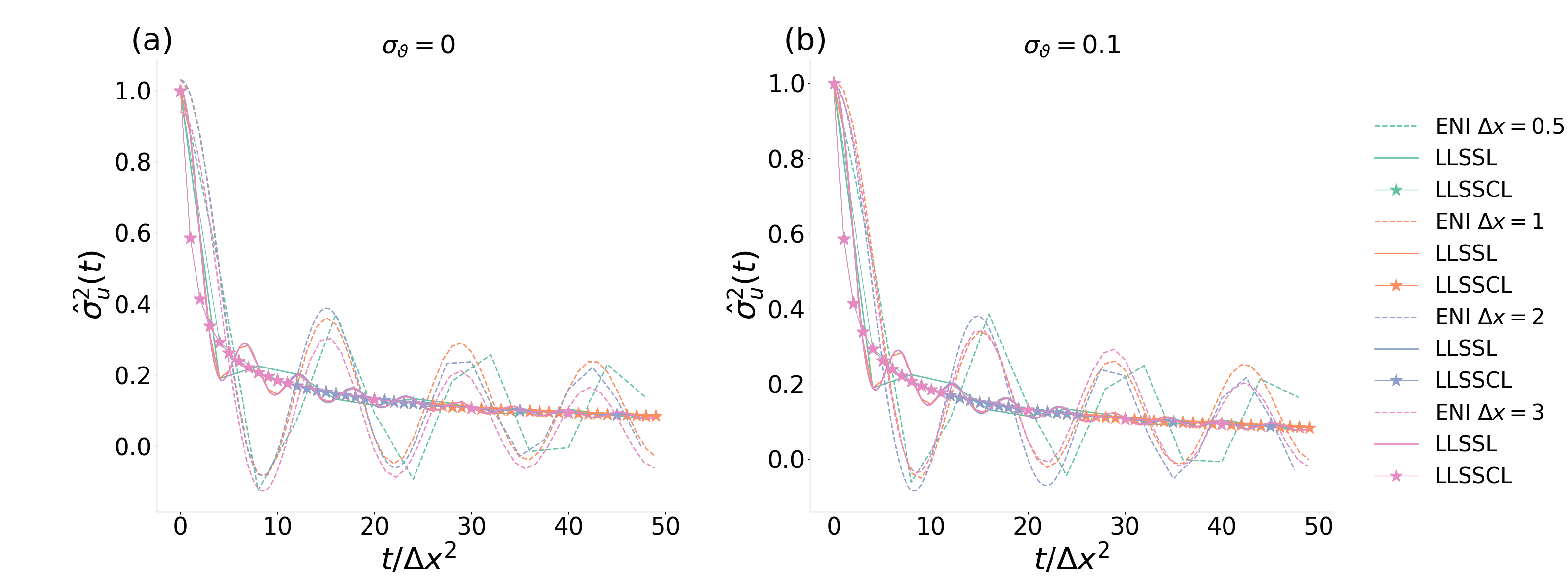}
		\caption{\textbf{Scaled variance of the amplitude fluctuations as a function of rescaled time $t / \Delta x ^2$ for different values of lattice discretization $\Delta x$.} Dashed lines are the exact numerical integration (ENI) of Schr\"{o}dinger equation, solid lines are the analytical prediction by taking the linearized large system-size limit (LLSSL), and the solid lines with stars are the approximation by taking the linearized large system-size and continuum limit (LLSSCL). The initial standard deviation of the amplitude is $\sigma_u = 0.05$. }
    \label{fig:scaled2_amplitude_variance}
	\end{figure}

We also observe that for various discretization values, $\Delta x$ the time-dependent variance curves by the exact numerical integration (ENI) collapse reasonably well to a scaling function which is a function of $t/(\Delta x)^2$ [Fig.~\ref{fig:scaled2_amplitude_variance}]. This natural scaling is also suggested by the results of the linear approximation (LLSSL)  [Eq.~\myref{scaled_variance}], but as pointed out above, the scaled LLSSL variance curves are distinct from the ENI ones, due to the strong non-linear effects.

\begin{figure}[H]
	\centering
		\includegraphics[width=1.0\textwidth]{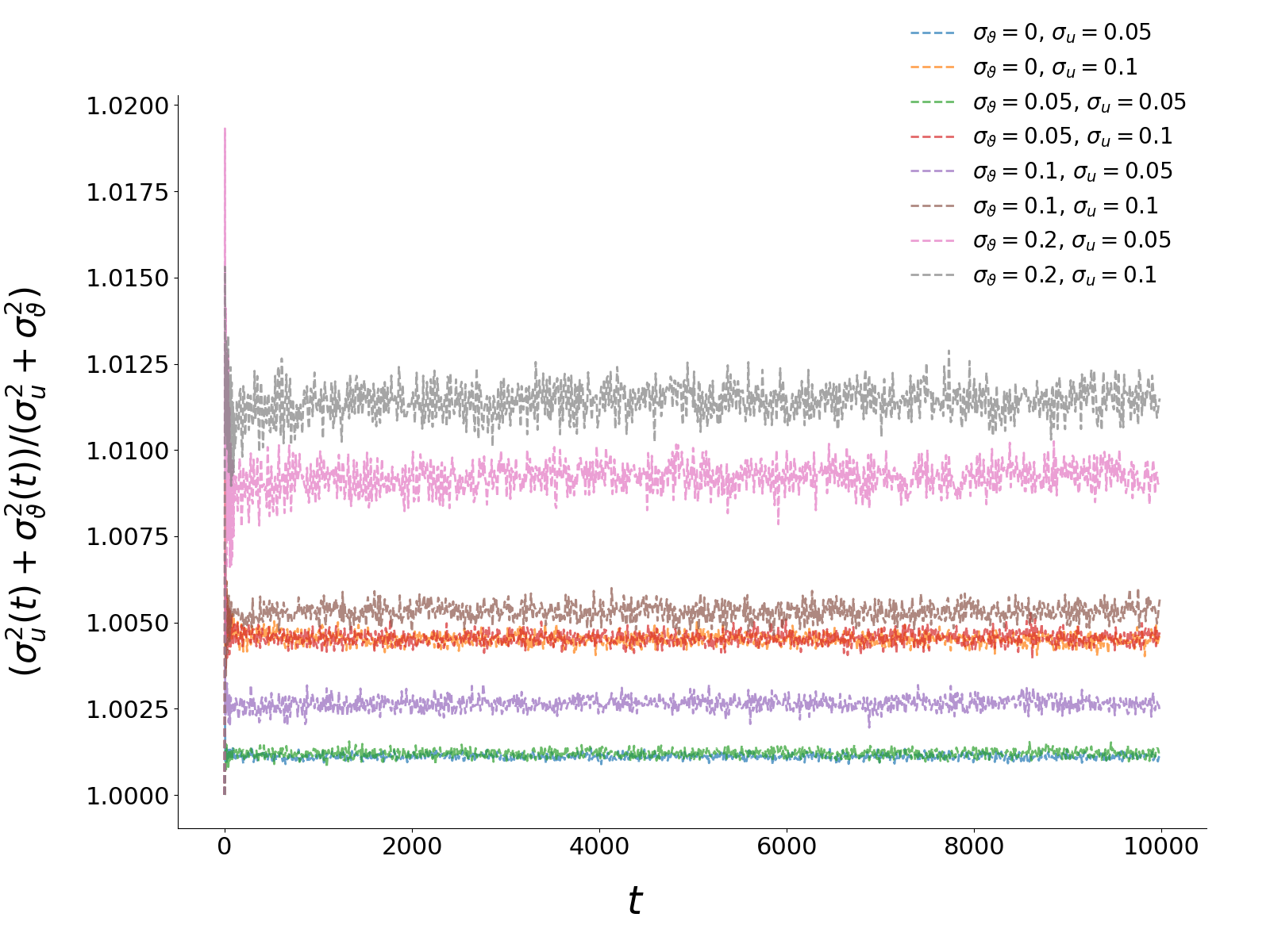}
		\caption{\textbf{The sum of the variances of the amplitude and phase fluctuations.}  The results are from exact numerical integration (ENI) of Schr\"{o}dinger equation.}
    \label{fig:sum_of_variances}
	\end{figure}

\begin{figure}[H]
	\centering
		\includegraphics[width=0.9\textwidth]{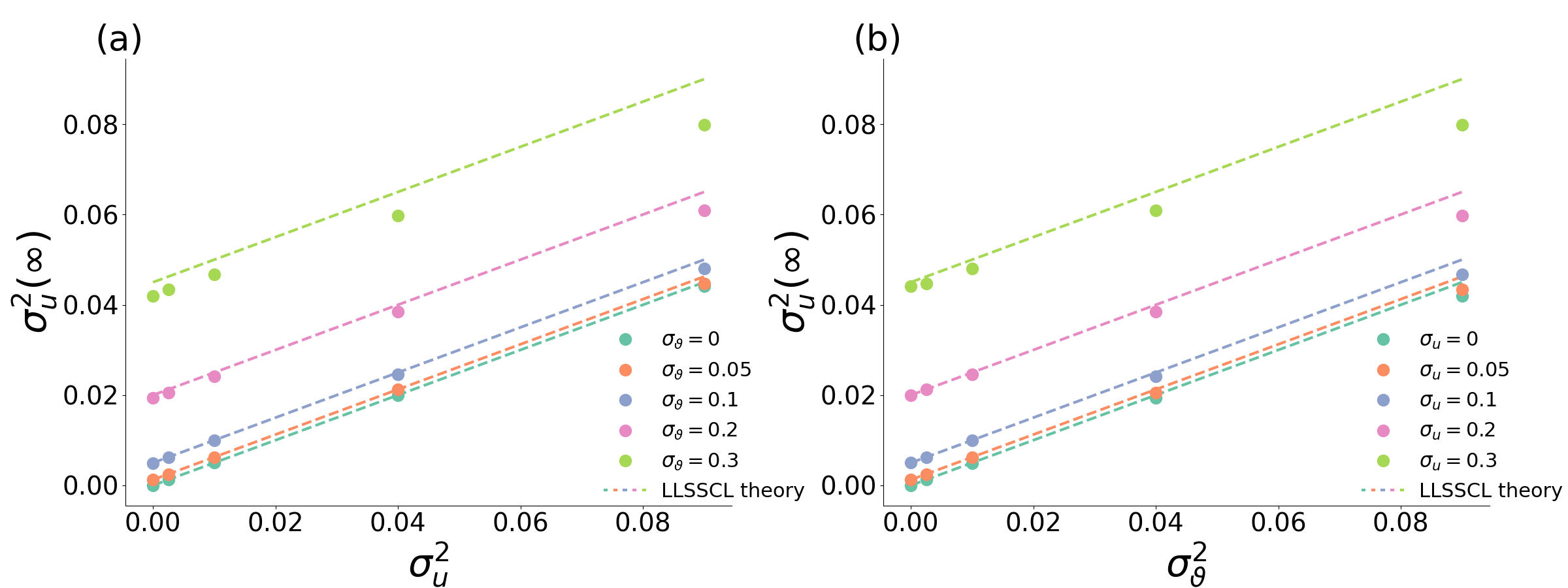}
		\caption{\textbf{Steady-state variance of the amplitude fluctuations $\sigma_u^2(\infty)$ as a function of the initial amplitude variance
        $\sigma_u^2$ and the initial phase variance $\sigma_{\vartheta}^2$.} The dashed lines are plotted according to Eq.~\myref{eq:asymptotic_variance}.}
    \label{fig:asymptotic_variance}
	\end{figure}

We also compared the results of the linear approximation for the sum of the time-dependent amplitude and phase variances [Eqs.~\myref{eq:sum_of_variances} and \myref{eq:sum_of_variances_scaled}, i.e., being a constant] and for the asymptotic steady-state variances as a function of the initial values  [Eq.~\myref{eq:asymptotic_variance}] with the ENI results
in Figs.~\ref{fig:sum_of_variances} and \ref{fig:asymptotic_variance}, respectively.
They both show reasonable agreement with the ENI data in the regime of small initial fluctuations. In particular, the relative errors of the predictions of the linear approximation are less than 2\% even for the largest initial variance values, and progressively become smaller for smaller values of initial variances, e.g., 0.1\% for the smallest one [Figs.~\ref{fig:sum_of_variances}].

\subsection{Single-site temporal correlation functions (autocorrelations)}

\begin{figure}[H]
	\centering
		\includegraphics[width=1.0\textwidth]{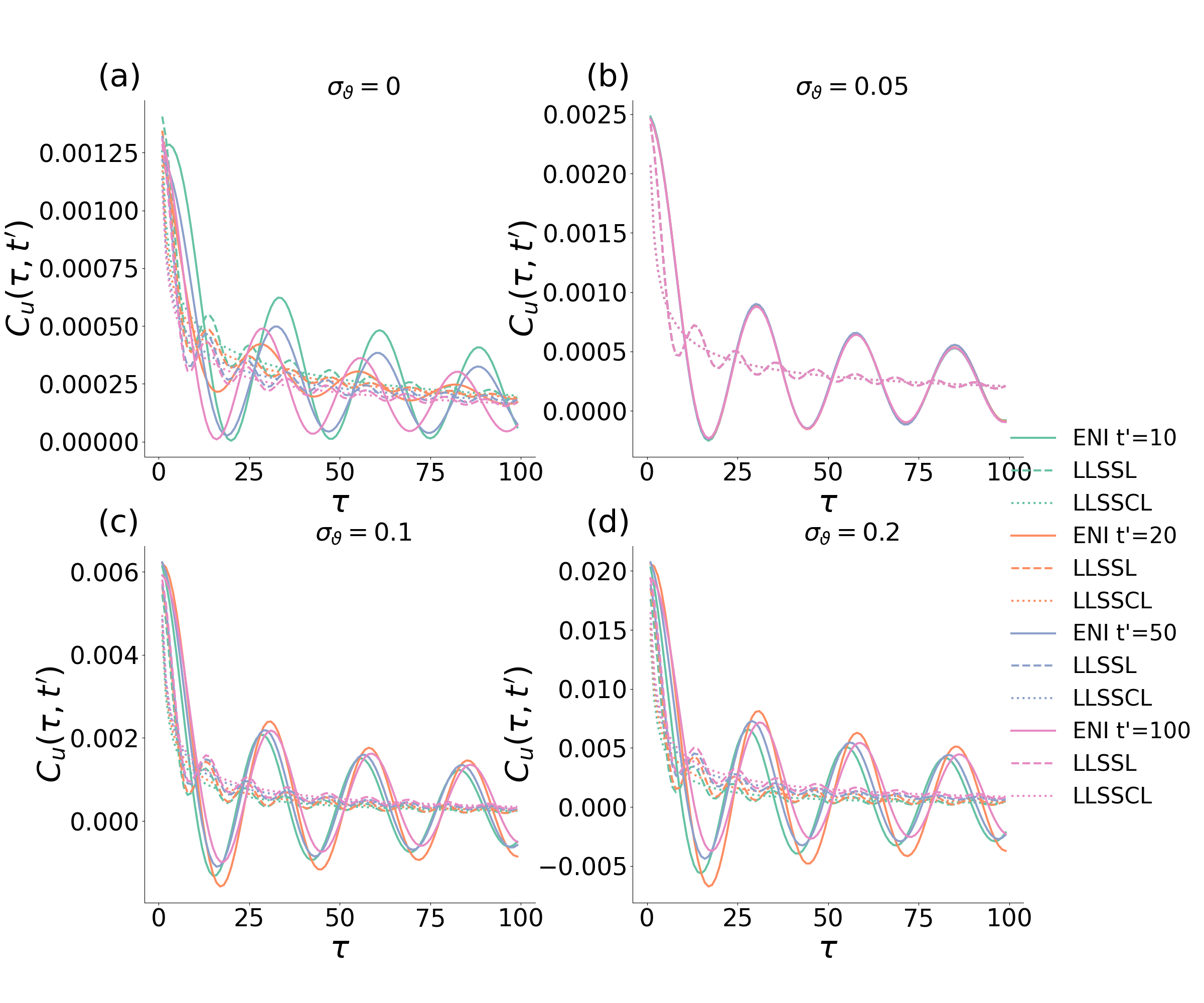}
		\caption{\textbf{Autocorrelation function of the amplitude fluctuations with time separation $\tau$ for different times. }  The initial standard deviation of the amplitude is set as $\sigma_u = 0.05$. In each subfigure, the initial standard deviation of the phase $\sigma_{\vartheta}$ is the same. Solid lines are the exact numerical integration (ENI) of Schr\"{o}dinger equation, dashed lines are the analytical prediction by taking the linearized large system-size limit (LLSSL), and the dotted lines are the approximation by taking the linearized large system-size and continuum limit (LLSSCL).}
  \label{fig:autocorr}
	\end{figure}

For the autocorrelation function of the amplitude and phase fluctuations, the comparison of the results of the linear approximation and the ENI results show agreements in some fundamental aspects, but also some key quantitative differences [Fig.~\ref{fig:autocorr}]. In particular, the autocorrelation functions asymptotically become time-homogeneous, e.g., $C_{u}(\tau,t')\equiv\langle u(\bm{x},t'+\tau) u(\bm{x},t')\rangle\simeq \hat{C}_{u}(\tau)$ for $t'\to \infty$. Further, in the regime of small fluctuations, for the special case $D_u$$=$$D_{\vartheta}$ ($\sigma^2_u$$=$$\sigma^2_{\vartheta}$), these autocorrelations are time-homogeneous for all times. The precise temporal scaling and period of the oscillatory decay by the ENI, however, differs from the results of linear approximation, analogous to those of the single-site variances. The asymptotic expansion of the error function in Eq.~\myref{autocorr_asymp} would suggest an oscillatory decay as a correction, $c'_1 \tau^{-1/2} + c'_2 \tau^{-1}\sin(\omega_{\max}\tau)$ [Eq.~\myref{autocorr_asymptotic}] with a period of $2\pi/\omega_{\max}=\frac{2(\Delta x)^2}{\alpha \pi} \approx 11$.
In contrast, the results of the exact numerical integration (ENI) indicate that the autocorrelation functions exhibit larger-amplitude oscillatory decay with a period of about 28.

\subsection{Persistence}

\begin{figure}[H]
	\centering
		\includegraphics[width=1.0\textwidth]{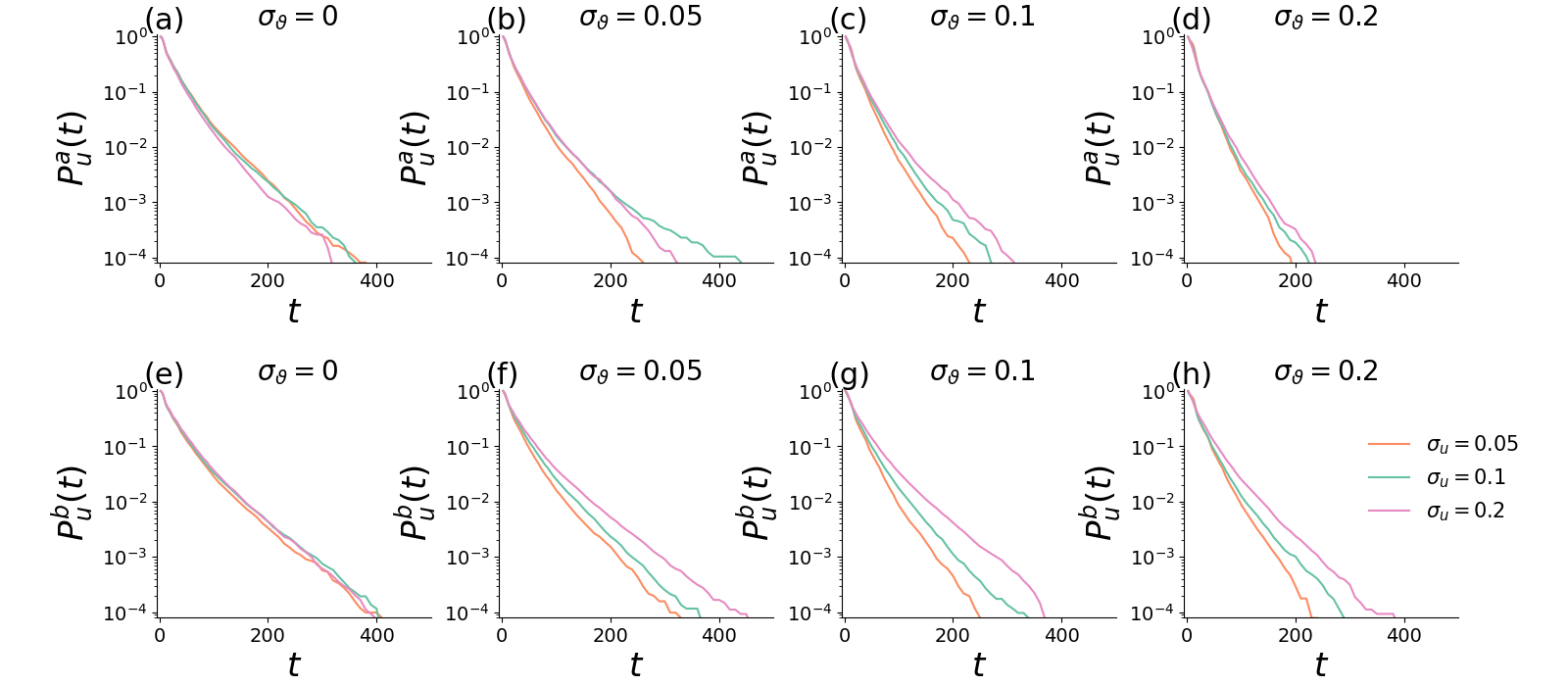}
		\caption{\textbf{Amplitude persistence probabilities $P_u^a(t)$ and $P_u^b(t)$.} In each subfigure, the initial standard deviation of the amplitude  $\sigma_{\vartheta}$ is the same. }
    \label{fig:amplitude_persistence}
	\end{figure}

The original equations [Eq.\myref{eq:sch_r_theta}]for the amplitude fluctuations are not symmetric above and below the mean, hence we must define corresponding persistence probabilities separately, $P^a_u(t)$ and $P^b_u(t)$, the probabilities that the local amplitude fluctuations have not changed sign up to time $t$, provided they were above and below the mean, respectively, at $t=0$. (Persistence probabilities for the phase fluctuations $P^a_{\vartheta}(t)$ and $P^b_{\vartheta}(t)$ are defined analogously.) In Fig.~\ref{fig:amplitude_persistence} we show results for the amplitude persistence $P^a_u(t)$ and $P^b_u(t)$.

\begin{figure}[H]
	\centering
		\includegraphics[width=1.0\textwidth]{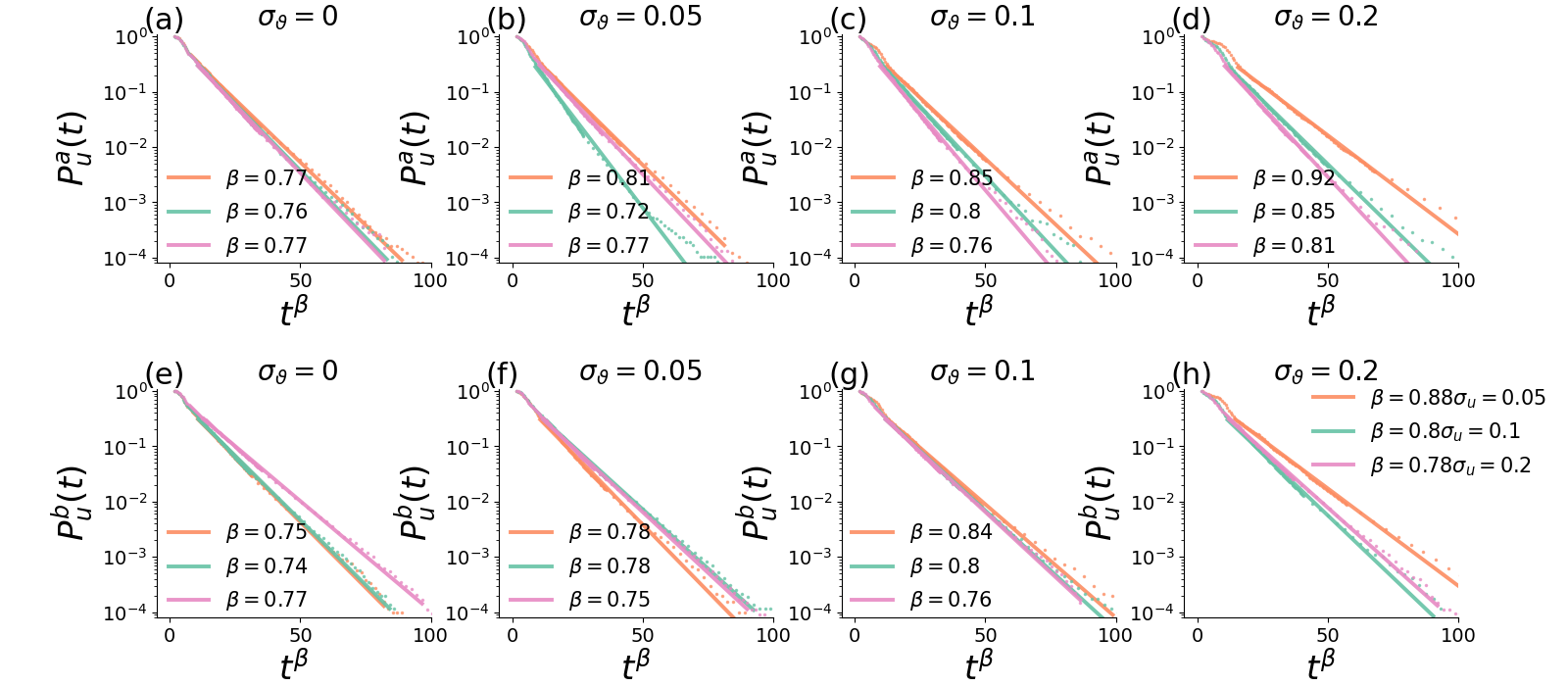}
		\caption{\textbf{Amplitude persistence probability $P_u^a(t)$ and $P_u^b(t)$ vs. $t^{\beta}$.} In each subfigure, the initial standard deviation of the amplitude  $\sigma_{\vartheta}$ is the same.}
    \label{fig:amplitude_persistence_stretched}
	\end{figure}

Using a nonlinear fit for Eq.\myref{eq:persistence_stretched_exp} we also show the scaled data ($P^a_u(t)$ and $P^b_u(t)$ vs. $t^{\beta}$) in Fig~\ref{fig:amplitude_persistence_stretched}, suggesting that the persistence probability for quantum fluctuations exhibit a stretched exponential tail ($\beta<1$). This is in line with the theoretical suggestions \cite{Newman1999,Feldheim_arxiv2021} that stationary (time-homogeneous) Gaussian processes with sufficiently rapidly decaying autocorrelation functions will exhibit stretched exponential persistence tails.

\begin{figure}[H]
	\centering
		\includegraphics[width=0.8\textwidth]{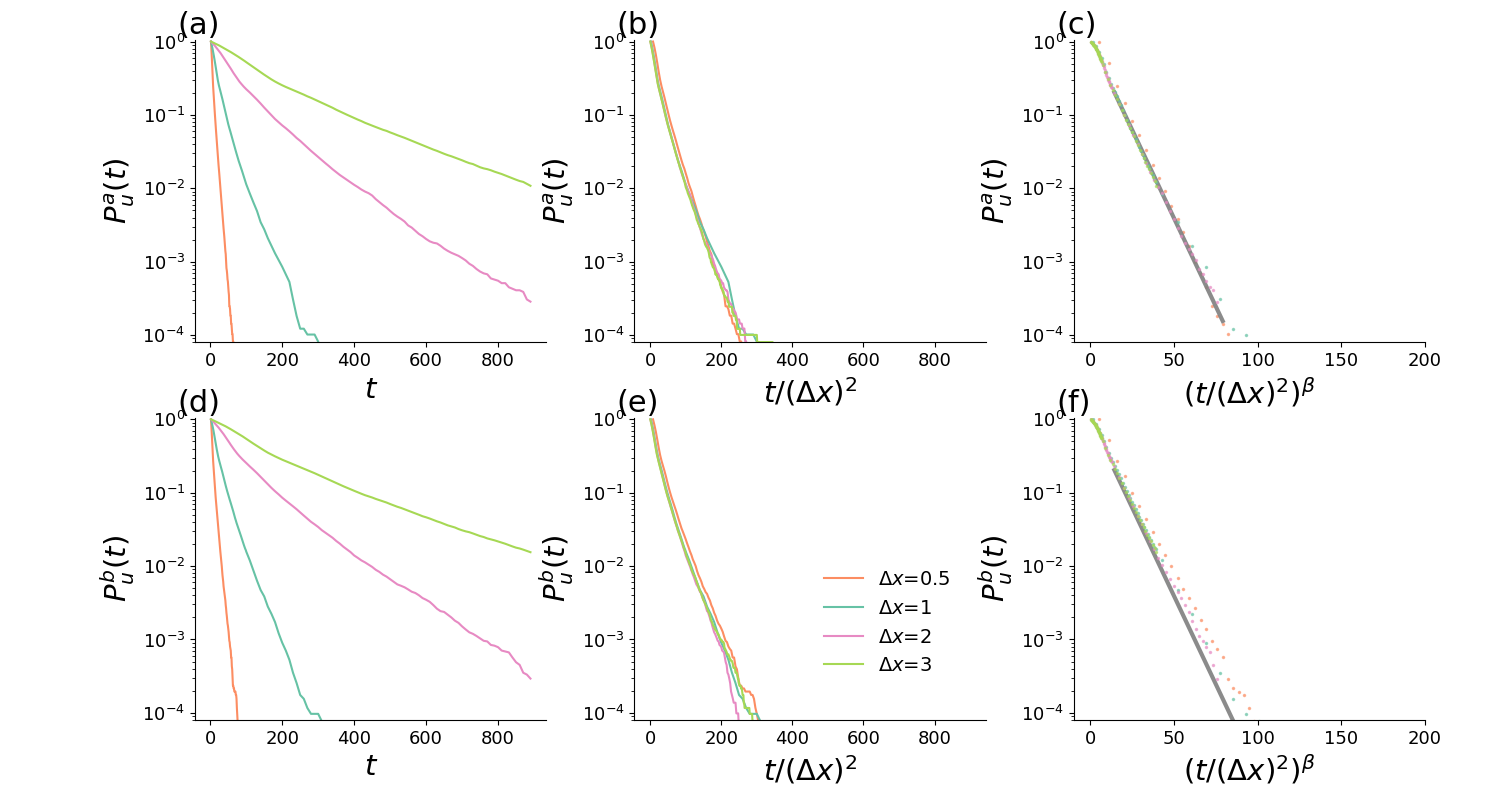}
		\caption{\textbf{Amplitude persistence probability $P_u^a(t)$ and $P_u^b(t)$ for different values of the lattice discretization $\Delta x$ and the corresponding scaling. } $N=1000$, $M=100$ realizations. $\sigma_{u}=0.05$ and $\sigma_{\vartheta}=0.05$.}
    \label{fig:amplitude_persistence_dx}
	\end{figure}

As one can anticipate from Schr\"odinger's equation [Eq.~\ref{eq:se_discretized}] with spatial discretization $\Delta x$, and from the form of the local time-dependent variances [Eq.~\myref{corr_spatial_discrete}], for various discretization values $\Delta x$ the persistence probability curves $P_u^a(t)$ and $P_u^b(t)$ collapse on a single scaling function which is a function of $t/(\Delta x)^2$ [Fig.~\ref{fig:amplitude_persistence_dx}].

\section{Higher dimensions}

\begin{figure}[H]
	\centering
		\includegraphics[width=1.0\textwidth]{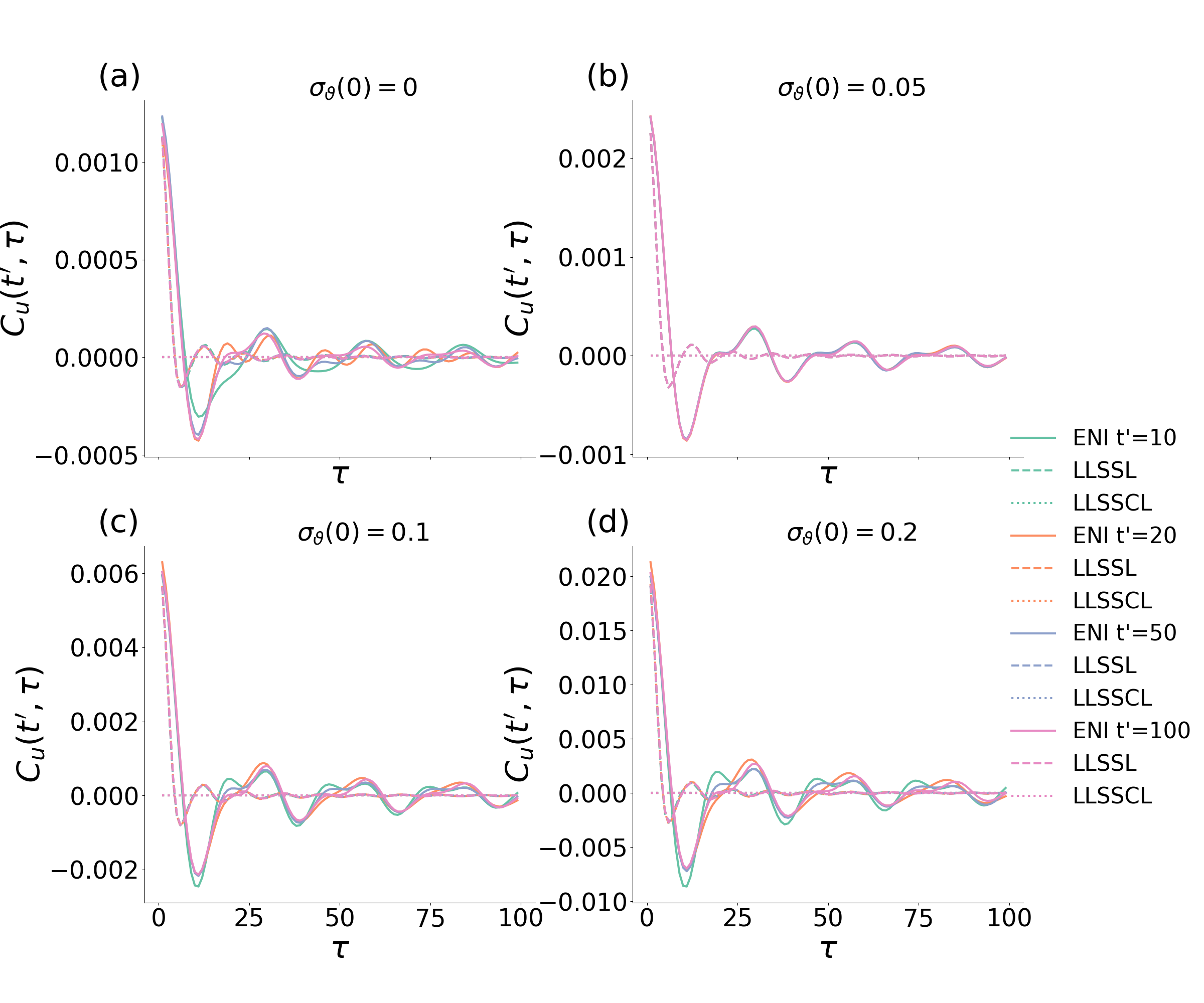}
		\caption{\textbf{Autocorrelation function of the amplitude fluctuations with time separation $\tau$ for different times in $d$$=$$2$.}  The initial standard deviation of the amplitude is set as $\sigma_u = 0.05$. In each subfigure, the initial standard deviation of the phase $\sigma_{\vartheta}$ is the same. Solid lines are the exact numerical integration (ENI) of Schr\"{o}dinger equation, dashed lines are the analytical prediction by taking the linearized large system-size limit (LLSSL), and the dotted lines are the approximation by taking the linearized large system-size and continuum limit (LLSSCL).}
  \label{fig:2D_amplitude_autocorr_func}
	\end{figure}

\begin{figure}[H]
	\centering
		\includegraphics[width=1.0\textwidth]{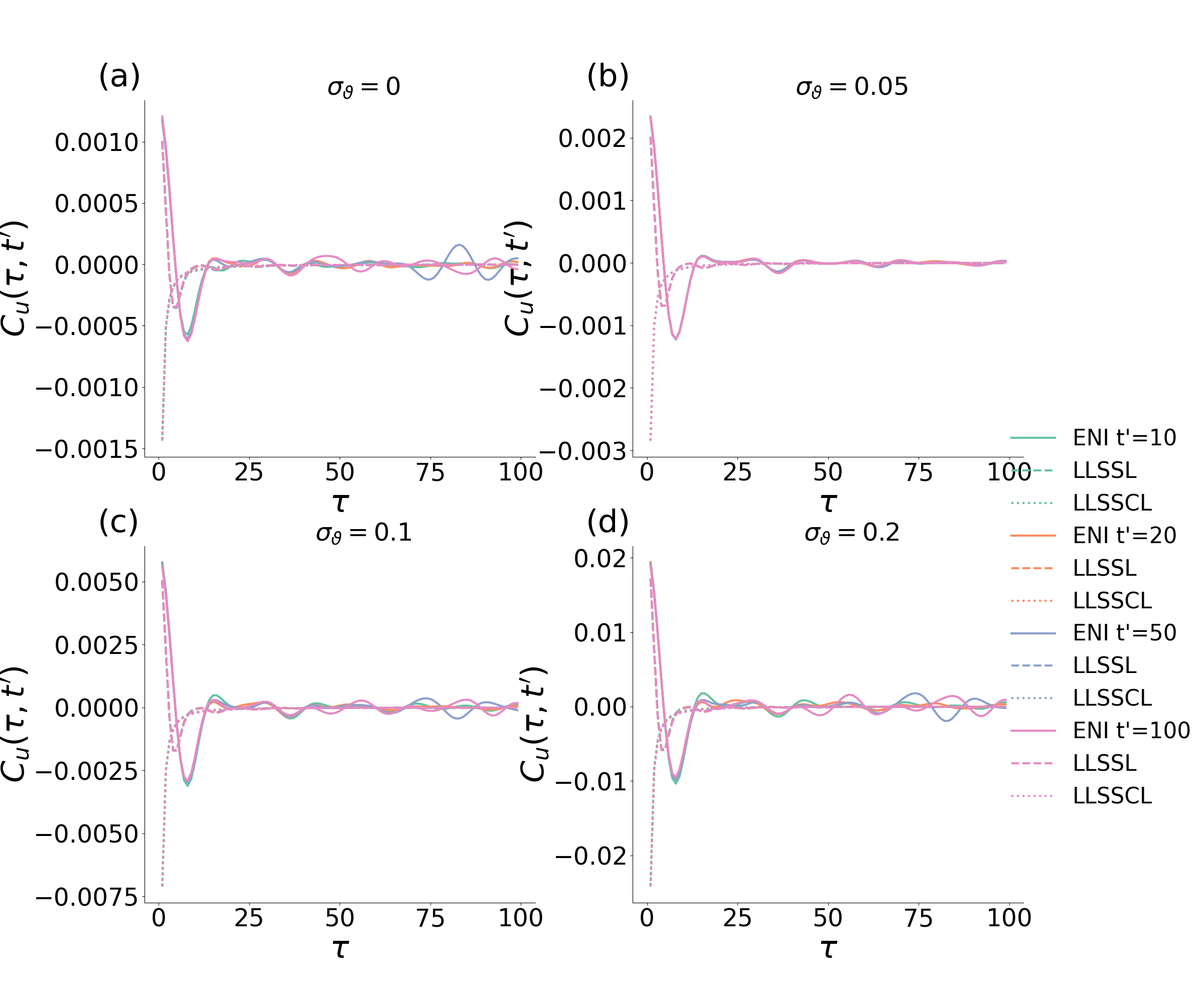}
		\caption{\textbf{Autocorrelation function of the amplitude fluctuations with time separation $\tau$ for different times in $d$$=$$3$.} The initial standard deviation of the amplitude is set as $\sigma_u = 0.05$. In each subfigure, the initial standard deviation of the phase $\sigma_{\vartheta}$ is the same. Solid lines are the exact numerical integration (ENI) of Schr\"{o}dinger equation, dashed lines are the analytical prediction by taking the linearized large system-size limit (LLSSL), and the dotted lines are the approximation by taking the linearized large system-size and continuum limit (LLSSCL).}
  \label{fig:3D_amplitude_autocorr_func}
	\end{figure}

While the predictions of the linear approximation still work well for the asymptotic steady-state variance for small amplitude and phase fluctuations [Eq.~\myref{eq:asymptotic_variance}] (see Supplemental Material/III.), the correlation functions in $d$$=$$2$ and $d$$=$$3$ exhibit a much more complex pattern. In particular, we observe strong deviations from regular sinusoidal oscillations and recurrent oscillations for long times, caused by nonlinearities becoming more relevant in higher dimensions. Here we show results for the autocorrelation functions in $d=2$ [Fig.~\myref{fig:2D_amplitude_autocorr_func}] and $d=3$ [Fig.~\myref{fig:3D_amplitude_autocorr_func}]. These observations are consistent with recent studies of turbulence in $d$$=$$2$ and $d$$=$$3$ equivalent fluid flow (with fluid density $\rho$$=$$R^2$ and velocity $\bm{v}\propto\nabla\vartheta$), generated by the free Schr\"{o}dinger equation \cite{Chiueh_2011}.

\begin{figure}[H]
	\centering
		\includegraphics[width=1.0\textwidth]{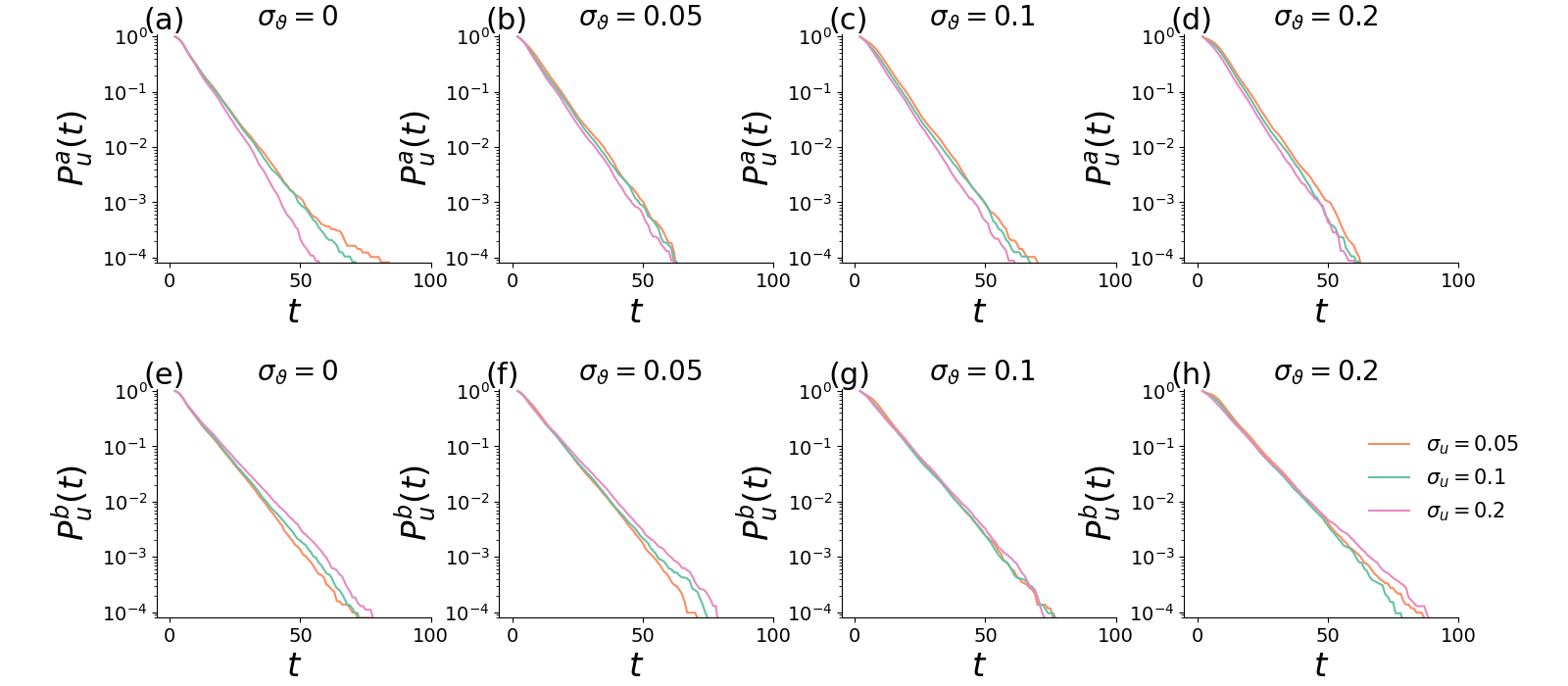}
		\caption{\textbf{Amplitude persistence probability $P_u^a(t)$ and $P_u^b(t)$ vs. $t$ in $d$$=$$2$.} In each subfigure, the initial standard deviation of the phase $\sigma_{\vartheta}$ is the same.}
  \label{fig_2D_amplitude_persistence}
	\end{figure}

\begin{figure}[H]
	\centering
		\includegraphics[width=1.0\textwidth]{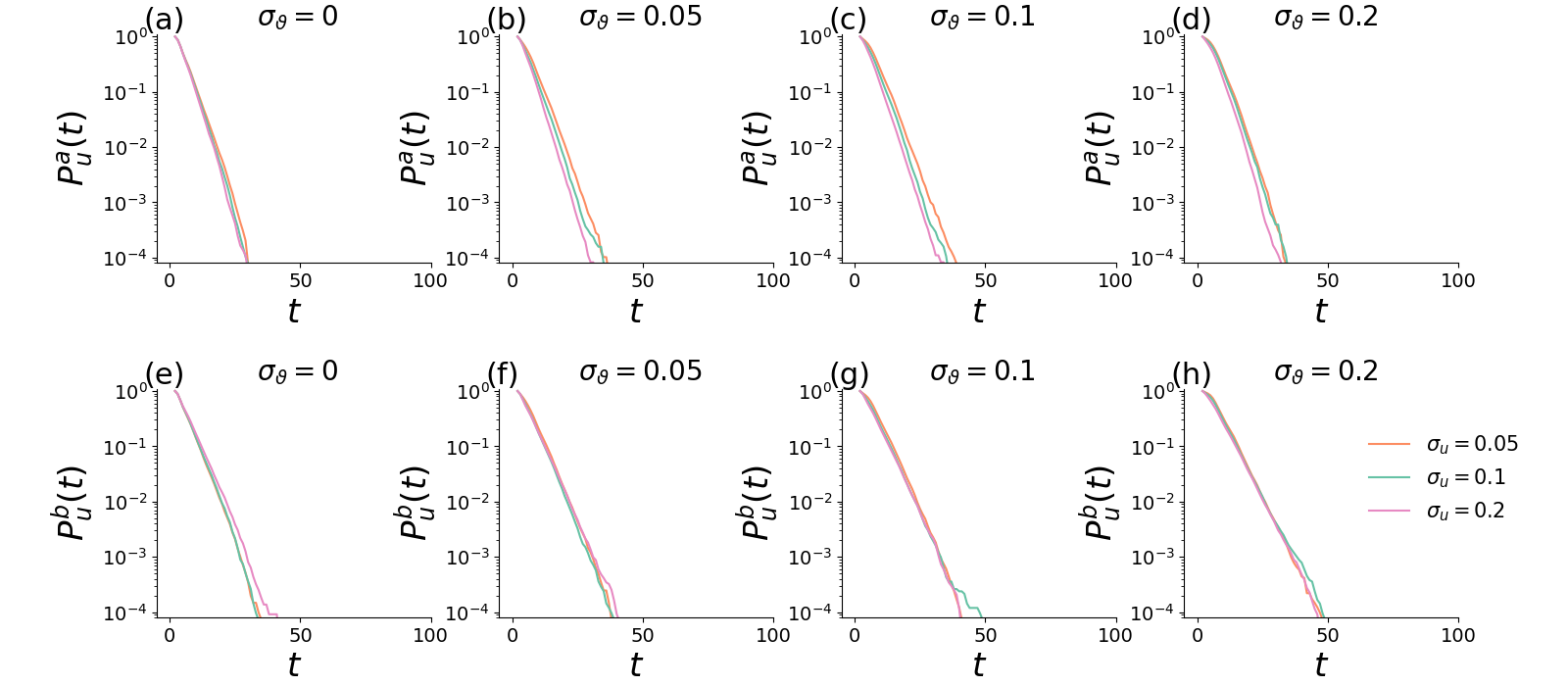}
		\caption{\textbf{Amplitude persistence probability $P_u^a(t)$ and $P_u^b(t)$ vs. $t$ in $d$$=$$3$.} In each subfigure, the initial standard deviation of the phase $\sigma_{\vartheta}$ is the same.}
  \label{fig:3D_amplitude_persistence}
	\end{figure}

Nevertheless, despite the strong quantitative and qualitative effects of the nonlinear terms in higher dimensions, it appears that the autocorrelation functions become approximately stationary (time-homogeneous) and decay sufficiently fast (even within recurring turbulent periods) so that the persistence probability distributions exhibit exponential tails in both $d$$=$$2$ [Fig.~\ref{fig_2D_amplitude_persistence}] and $d$$=$$3$ [Fig.~\ref{fig:3D_amplitude_persistence}]. We also show the persistence probabilities for a few other values of $\Delta x$, and the corresponding scaling behavior with $t/(\Delta x)^2$ in Supplemental Material/IV.

We note that despite the strong nonliner effects giving rise to turbulent flows in higher dimensions, in the regime of small fluctuations, the special symmetry seems to hold for the autocorrelation functions: for $D_u$$=$$D_{\vartheta}$ ($\sigma^2_u$$=$$\sigma^2_{\vartheta}$) the autocorrelations are time-homogeneous for all times (as suggested by the linear approximation Eq.~\myref{corr_temporal_discrete}) [Fig.~\myref{fig:2D_amplitude_autocorr_func}(b)] and [Fig.~\myref{fig:3D_amplitude_autocorr_func}(b)].

\section{A physical realization of the initial noise with an effective lengthscale cutoff}

So far we have assumed that the initial noise is delta-correlated, i.e., uncorrelated beyond the discretization length $\Delta x$. In this case, the choice of $\Delta x$ is an artifact of the spatial discretization scheme and has no particular connection to an underlying physical lengthscale.

\begin{figure}[H]
	\centering
		\includegraphics[width=0.9\textwidth]{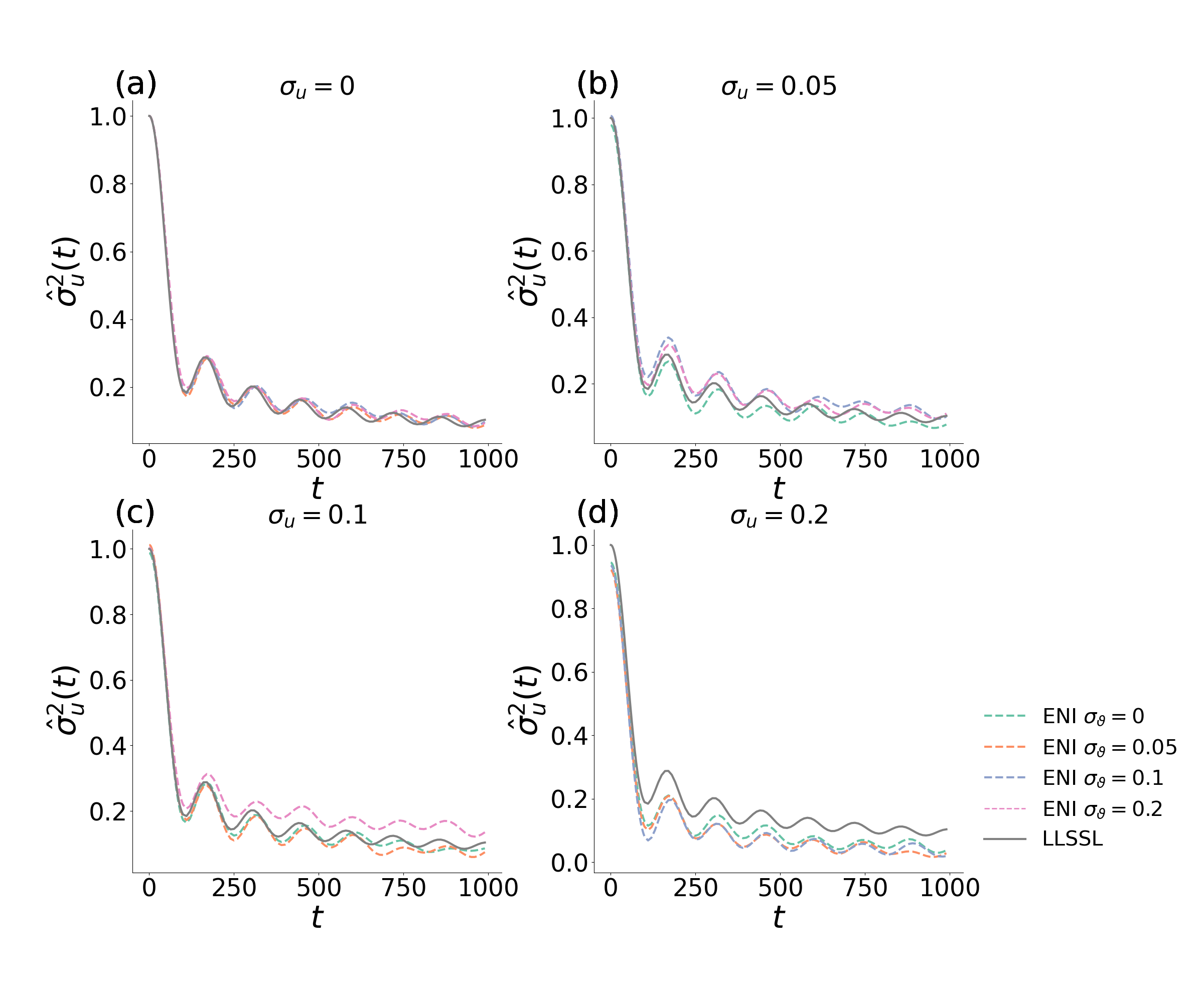}
 		\caption{\textbf{Scaled variance of the amplitude fluctuations as a function of time with an effective initial noise lengthscale cutoff $a=5\Delta x$ ($\Delta x$$=$$1$) in $d$$=$$1$.}
   In each subfigure, the initial standard deviation $\sigma_u$ is the same. Dashed lines are the exact numerical integration (ENI) of Schr\"{o}dinger equation, and solid lines are the analytical prediction by taking the linearized large system-size limit (LLSSL). }
   \label{fig:scaled_u_corr_a_cutoff}
	\end{figure}

\begin{figure}[H]
	\centering
		\includegraphics[width=1.0\textwidth]{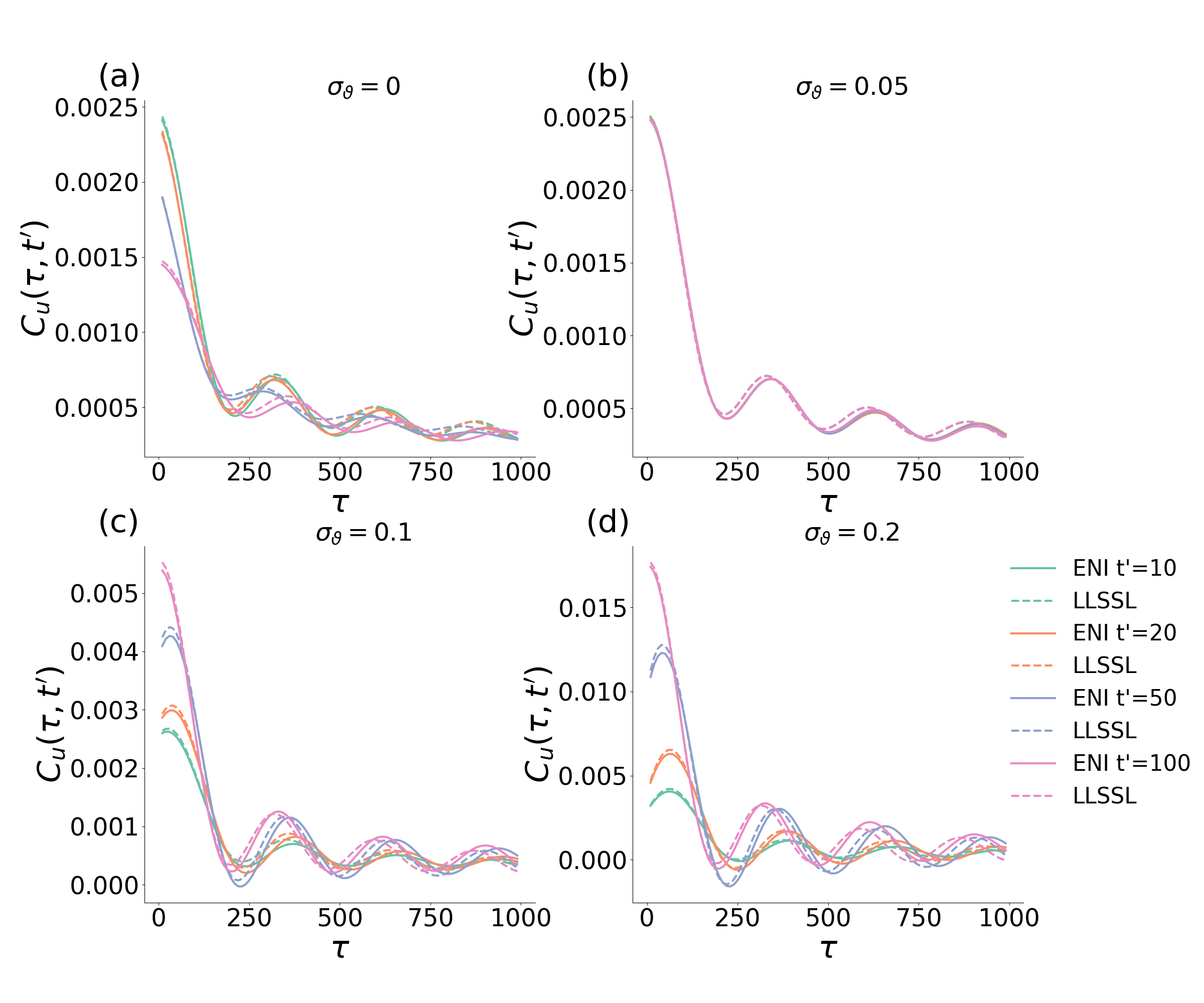}
		\caption{\textbf{Autocorrelation function of the amplitude fluctuations with time separation $\tau$ for different times with an effective initial noise lengthscale cutoff $a=5\Delta x$ ($\Delta x$$=$$1$) in $d$$=$$1$.} The initial standard deviation of the amplitude is set as $\sigma_u = 0.05$. In each subfigure, the initial standard deviation of the phase $\sigma_{\vartheta}$ is the same. Solid lines are the exact numerical integration (ENI) of Schr\"{o}dinger equation, dashed lines are the analytical prediction by taking the linearized large system-size limit (LLSSL).}
  \label{fig:u_autocorr_a_cutoff}
	\end{figure}

Now we consider the case where the lengthscale associated with the initial noise distribution is the result of some underlying physical process with a corresponding effective lengthscale $a$ ($\Delta x\ll a\ll L$). As illustrated in Appendix~\ref{appendix_D}, one can generate such initial noise by imposing a hard cutoff in momentum space, i.e., all modes have zero amplitudes for $k>k_{\rm max}=\pi/a$. Consequently, the real-space initial noise structure becomes
$\langle \eta(\bm{x})\rangle=0$, $\langle \xi(\bm{x})\rangle=0$, and
\begin{equation}
\begin{split}
& \langle \eta(\bm{x})\eta(\bm{x}')\rangle = D_u \delta_{(a)}(\bm{x}-\bm{x}')
= D_u \prod_{j=1}^{d} \delta_{(a)}(x_j-x'_j)
= D_u \prod_{j=1}^{d} \frac{\sin(\pi(x_j-x'_j)/a)}{\pi (x_j-x'_j)}
\\
& \langle \xi(\bm{x})\xi(\bm{x}')\rangle =   D_{\vartheta} \delta_{(a)}(\bm{x}-\bm{x}')
= D_{\vartheta} \prod_{j=1}^{d} \delta_{(a)}(x_j-x'_j)
= D_{\vartheta} \prod_{j=1}^{d} \frac{\sin(\pi(x_j-x'_j)/a)}{\pi (x_j-x'_j)}  \\
& \langle \eta(\bm{x})\xi(\bm{x}')\rangle = 0  \;\;\;,
\label{eq:noise_correlation_with_cutoff}
\end{split}
\end{equation}
where
\begin{equation}
\delta_{(a)}(x)\equiv \frac{\sin(\pi x/a)}{\pi x}
\;\;\;.
\end{equation}
Then the initial local variances become
$\sigma_u^2 = \langle \eta^2(\bm{x})\rangle = D_u/a^d$
and
$\sigma_{\vartheta}^2 = \langle\xi^2(\bm{x})\rangle = D_{\vartheta}/a^d$.
With the above ``renormalized" initial noise, one can realize that in the subsequent calculations, the only changes will be the replacement of
$\frac{\delta_{\bm{x},\bm{x}'}}{(\Delta x)^d}$ (or $\delta(\bm{x}-\bm{x}')$) with $\delta_{a}(\bm{x}-\bm{x}')$ and $\Delta x$ with $a$ in the expressions involving the error functions. For example, for the equal-time amplitude correlations we have,
\begin{equation}
\langle u(\bm{x},t) u(\bm{x}',t)  \rangle  \simeq
\frac{D_{u} + D_{\vartheta}}{2}  \delta_{a}(\bm{x}-\bm{x}')
+  \frac{D_{u} - D_{\vartheta}}{2} \frac{1}{(8\pi\alpha t)^{d/2}}
   \Re \left( \prod_{j=1}^{d} F(x_j-x'_j,t;a) \right)
\;\;\;,
\end{equation}
and for its special case, the single-site variance,
\begin{equation}
\sigma^2_u(t) = \langle u^2(\bm{x},t)  \rangle  \simeq
\frac{D_{u} + D_{\vartheta}}{2}  \frac{1}{a^d}
+  \frac{D_{u} - D_{\vartheta}}{2} \frac{1}{(8\pi\alpha t)^{d/2}}
   \Re \left( F^d(0,t;a) \right)
\;\;\;.
\end{equation}
Similarly, one can work out the autocorrelation function, e.g., for the amplitude fluctuations,
\begin{equation}
\begin{split}
& \langle u(\bm{x},t) u(\bm{x},t')  \rangle  \simeq \\
& \frac{D_{u} + D_{\vartheta}}{2} \frac{1}{(4\pi\alpha |t-t'|)^{d/2}}   \Re \left( H^d(t-t';a) \right)
+
\frac{D_{u} - D_{\vartheta}}{2} \frac{1}{(4\pi\alpha (t+t'))^{d/2}}   \Re \left( H^d(t+t';a) \right)
\;\;\;,
\end{split}
\end{equation}
and specifically, in the asymptotic long-time limit, $t=t'+\tau$ with $\tau$ fixed and $t'\to \infty$,
\begin{equation}
C_{u}(\tau,t')\equiv\langle u(\bm{x},t'+\tau) u(\bm{x},t')\rangle\simeq \hat{C}_{u}(\tau) \simeq
\frac{D_{u} + D_{\vartheta}}{2} \frac{1}{(4\pi\alpha |\tau|)^{d/2}}   \Re \left( H^d(\tau;a) \right)
\;\;\;.
\end{equation}
As the above results from the linear approximation indicate, the single-site fluctuations and the autocorrelation functions are now governed by the ``physical" effective lengthscale $a$, separate from the numerical discretization $\Delta x$ (which can be taken to $0$). Since the correlation functions are now independent of $\Delta x$, this renormalization also eliminates the progressively increasing frequency in the oscillatory decays. Further, since the nonlinear terms that we neglected are rather sensitive to the wave numbers present initially, e.g., $(\nabla r) (\nabla \vartheta)$ in Eq.~\myref{eq:sch_r_theta} or its corresponding representation $\propto\sum_{\bm{k}',\bm{k}''} \delta_{\bm{k},\bm{k}'+\bm{k}''} (\bm{k}'\cdot\bm{k}'')\tilde{u}_{\bm{k}'} \tilde{\vartheta}_{\bm{k}''}$ in momentum space, we can expect the linear approximation to work better for small fluctuations in a perturbative sense. While such couplings do generate higher wave-number modes with time which were not initially present, for small initial densities their effect is negligible, especially for early times. Here we show the scaled single-site variance [Fig.~\myref{fig:scaled_u_corr_a_cutoff}] and the autocorrelations [Fig.~\myref{fig:u_autocorr_a_cutoff}] for the amplitude fluctuations, indicating a reasonable agreement with the frequency of oscillations (governed by the physical lengthscale), $2\alpha\left(\frac{\pi}{a}\right)^2$ and $\alpha\left(\frac{\pi}{a}\right)^2$, respectively (the corresponding periods are $\sim$$137$ and $\sim$$275$, respectively).

\begin{figure}[H]
	\centering
		\includegraphics[width=1.0\textwidth]{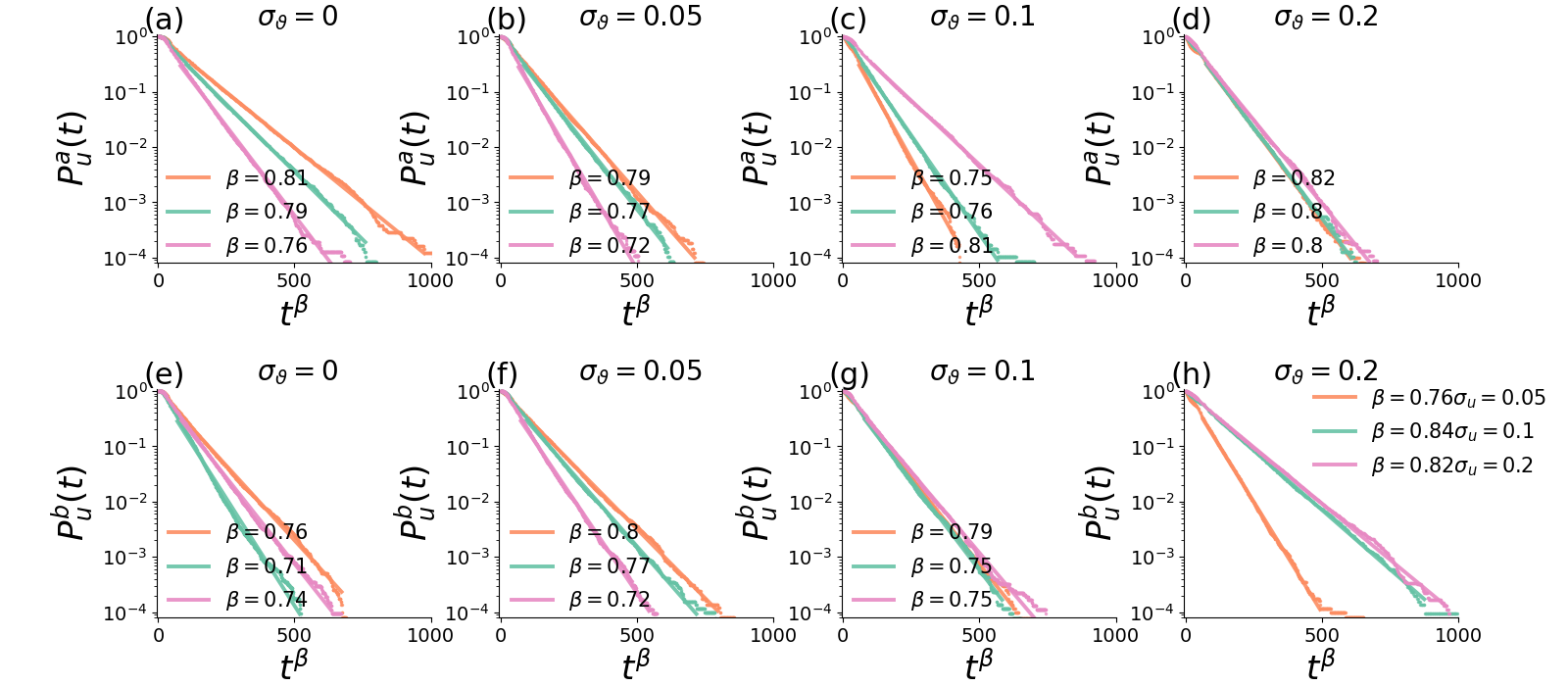}
		\caption{\textbf{Scaled amplitude persistence probability $P_u^a(t)$ and $P_u^b(t)$ vs. $t^{\beta}$ for an effective initial noise cutoff $a=5\Delta x$ ($\Delta x$$=$$1$).} In each subfigure, the initial standard deviation of the amplitude  $\sigma_{\vartheta}$ is the same.}
  \label{fig:u_persistence_a_cutoff_stretched}
	\end{figure}

Since the autocorrelation function (at least for small initial fluctuations) is asymptotically time-homogeneous and its magnitude decays sufficiently fast [approximately given by Eq.~\myref{autocorr_asymptotic_cutoff_a}], we anticipate the persistence probabilities to follow stretched exponentials, as confirmed by the numerical results shown in Fig.~\myref{fig:u_persistence_a_cutoff_stretched}.

\section{Summary and Outlook}

As it is well-known, the imaginary unit $i$ makes all the difference between the diffusion equation and the free-particle Schr\"odinger's equation: the former giving rise to classical relaxation (i.e., the linear combination of exponentially decaying modes) while the latter giving rise to the superposition of waves. Here we studied the amplitude and phase persistence probability in free-particle quantum diffusion, governed by Schr\"odinger's equation.

Employing the Crank-Nicholson method, we numerically integrated the free-particle Schr\"odinger equation from locally random uncorrelated initial conditions, with small Gaussian amplitude fluctuations about the homogeneous density, and small phase fluctuations. In the limit of small initial fluctuations, the fluctuations of the (probability) density are proportional to those of the amplitude of the wave function. Our exact numerical integration (ENI) results indicate that both the amplitude and phase persistence probability decay in a stretched exponential fashion in $d$$=$$1$, while they exhibit exponential tails in $d$$=$$2$ and $d$$=$$3$. These are consistent with sufficiently fast-decaying autocorrelation functions of asymptotically stationary (time-homogeneous) processes \cite{Newell_1962}, and supported by our numerical results.

To gain some insight, we studied the linearized coupled oscillations of the amplitude and phase fluctuations. In $d$$=$$1$ there is a good quantitative agreement between the trend of the decay of the local site variances and the autocorrelation functions, while the precise period and amplitude of the oscillatory behavior are not captured correctly because of the effects of the wave-number-dependent non-linear terms. Note the agreement between linear approximation and the ENI can be improved (at least for small initial variances) when the high wave-number modes in the amplitude and phase fluctuations are eliminated initially, which corresponds to a spatially-correlated initial noise distribution, controlled by a finite-lengthscale cutoff [Eq.~\myref{eq:noise_correlation_with_cutoff}].

In $d$$=$$2$ and $d$$=$$3$, the linear approximation misses irregular and recurrent oscillatory patterns governed by the non-linear terms, possibly as a result of turbulence in higher dimensions \cite{Chiueh_2011}.

The observed quantum fluctuations also exhibit some interesting properties in the studied small-initial-variance regime for the amplitude and phase fluctuations: for $D_u$$=$$D_{\vartheta}$ ($\sigma^2_u$$=$$\sigma^2_{\vartheta}$) (\textit{i}) the quantum system becomes {\em spatially correlation-free} at all times; (\textit{ii})  the single-site temporal correlation functions (the autocorrelations) are time-homogeneous for all times (otherwise they are time-homogeneous in the asymptotic long-time limit). The latter property (time-homogeneity) is the key underlying reason \cite{Newell_1962} (and the key difference with respect to classical diffusion \cite{Bray2013}) that the corresponding persistence probabilities are exponential like: stretched exponential in $d$$=$$1$ and approximately exponential in $d$$=$$2$ and $d$$=$$3$.

We note that the approximate forms of the amplitude and phase correlation functions presented in this paper are non-universal and are applicable only to the regime of small amplitude and phase fluctuations ($\sigma_u\ll 1$, $\sigma_{\vartheta}\ll 1$). For example, one can show that for a uniform initial phase distribution in $[0,2\pi)$, the asymptotic steady-state amplitude variance becomes independent of the initial size of the amplitude fluctuations,  $\sigma^2_u(\infty)\simeq 1/4$. Nevertheless, even in this case, the autocorrelation functions decay sufficiently fast and become approximately time-homogeneous, giving rise to exponential-like persistence tails \cite{Newell_1962}.

While in this current work we investigated amplitude and phase fluctuations and persistence in quantum diffusion on regular lattices, there are equally interesting and analogous questions that can be asked for quantum diffusion on disordered lattices \cite{Anderson_PR1958,Domany_PRB1983,Soukoulis_PRB1992,Wang_PRE2013} and random structures (e.g., small-world interconnects) \cite{Zhu_PRB2000,Kim_PRB2003,Mulken_PRE2007,Novotny_Procedia2014,Novotny_PRA2018}.
In particular, subject to spatially random initial conditions, what the relaxation and persistence properties of the amplitude and phase fluctuations are, and whether they can provide useful characterization and signatures \cite{Moro_ChemPhys2018} in the context of Anderson localization \cite{Anderson_PR1958}. Current efforts are under way in these directions \cite{Korniss_2024}.


\appendix

\section{Solving the linearized coupled equations for the amplitude and phase fluctuations}
\label{appendix_A}

Introducing the Fourier transforms
	\begin{equation}
	\begin{split}
		u(\bm{x}, t) & = \frac{1}{L^d} \sum _{\bm{k}}^{} \tilde{u}_{\bm{k}}(t) e^{i\bm{k \cdot x}} \\
	\vartheta(\bm{x}, t)  & = \frac{1}{L^d} \sum _{\bm{k}}^{} \tilde{\vartheta}_{\bm{k}}(t) e^{i\bm{k \cdot x}} 	
 \end{split} \;\;\;,
 \label{eq:fourier}
\end{equation}
where $L$ is the linear size of the lattice, from Eq.~\myref{eq:u_coupled}] we find
 \begin{equation}
	 \begin{split}
	 \partial_t \tilde{u}_{\bm{k}} & = \alpha \bm{k}^2 \tilde{\vartheta}_{\bm{k}}  \\
	 \partial_t \tilde{\vartheta}_{\bm{k}} & = - \alpha \bm{k}^2 \tilde{u}_{\bm{k}}
	 \end{split}   \;\;\;,
  \label{eq:u_k_coupled}
\end{equation}
or written in a matrix form,
\begin{equation}
\partial_t
\begin{pmatrix}
\tilde{u}_{\bm{k}} \\
\tilde{\vartheta}_{\bm{k}}
\end{pmatrix}
=
\begin{pmatrix}
0 & \alpha \bm{k}^2 \\
- \alpha \bm{k}^2 & 0
\end{pmatrix} \\
\begin{pmatrix}
\tilde{u}_{\bm{k}} \\
\tilde{\vartheta}_{\bm{k}}
\end{pmatrix} \;\;\;.
\label{eq:u_k_matrix}
\end{equation}
Note that here $\alpha=\frac{\hbar}{2m}$ is real and positive. The above simple system has two linearly independent solutions with oscillatory time dependence $e^{\pm i\omega_{\bm{k}}t}$, where $\omega_{\bm{k}}=\alpha \bm{k}^2 = \alpha \sum _{j=1}^{d} k_j^2$, as expected (i.e., the non-relativistic energy-momentum dispersion of a free particle). After finding the corresponding eigenvectors, the general solution can be written as
\begin{equation}
\begin{pmatrix}
\tilde{u}_{\bm{k}}(t) \\
\tilde{\vartheta}_{\bm{k}}(t)
\end{pmatrix}
=
A_{\bm{k}}
\begin{pmatrix}
1 \\
i
\end{pmatrix}  e^{i\omega_{\bm{k}}t}
+
B_{\bm{k}}
\begin{pmatrix}
1 \\
-i
\end{pmatrix}  e^{-i\omega_{\bm{k}}t}
\;\;\;,
\label{eq:u_k_general}
\end{equation}
where $A_{\bm{k}}$ and $B_{\bm{k}}$ are to be determined from the initial conditions $\tilde{u}_{\bm{k}}(0)$ and $\tilde{\vartheta}_{\bm{k}}(0)$,
\begin{equation}
\begin{split}
A_{\bm{k}} & = \frac{1}{2}(\tilde{u}_{\bm{k}}(0) - i\tilde{\vartheta}_{\bm{k}}(0)) \\
B_{\bm{k}} & = \frac{1}{2}(\tilde{u}_{\bm{k}}(0) + i\tilde{\vartheta}_{\bm{k}}(0))
\end{split} \;\;\;.
\end{equation}
Here, we consider uncorrelated Gaussian random variables for the initial amplitude and noise fluctuations, $\tilde{u}_{\bm{k}}(0)=\tilde{\eta}_{\bm{k}}$ and  $\tilde{\vartheta}_{\bm{k}}(0)=\tilde{\xi}_{\bm{k}}$. Hence, we can write the general solution for each mode of the amplitude and phase fluctuations
\begin{equation}
\begin{split}
\tilde{u}_{\bm{k}}(t) & = \frac{1}{2}( \tilde{\eta}_{\bm{k}} - i\tilde{\xi}_{\bm{k}} ) e^{i\omega_{\bm{k}}t} +
 \frac{1}{2}(\tilde{\eta}_{\bm{k}} + i\tilde{\xi}_{\bm{k}} ) e^{-i\omega_{\bm{k}}t} \\
\tilde{\vartheta}_{\bm{k}}(t) & = \frac{1}{2}(i\tilde{\eta}_{\bm{k}} + \tilde{\xi}_{\bm{k}} ) e^{i\omega_{\bm{k}}t} +
 \frac{1}{2}(-i\tilde{\eta}_{\bm{k}} + \tilde{\xi}_{\bm{k}} ) e^{-i\omega_{\bm{k}}t}
\end{split}   \;\;\;.
\label{eq:u_k_solutions}
\end{equation}

\section{Equal-time spatial correlation functions}
\label{appendix_B}

Here we show the key steps to obtain the amplitude-amplitude correlation function, the other two can be obtained analogously. For the spatially delta-correlated random initial variables (as defined in Sec.~\ref{sec:analytic}), their correlations for the Fourier components become
$\langle   \eta_{\bm{k}} \rangle = 0$,  $\langle   \xi_{\bm{k}} \rangle = 0$, $ \langle   \tilde{\eta}_{\bm{k}}\tilde{\eta}_{\bm{k}'}\rangle = D_u L^d \delta_{\bm{k}, -\bm{k}'}$,  $ \langle   \tilde{\xi}_{\bm{k}}\tilde{\xi}_{\bm{k}'}\rangle = D_{\vartheta} L^d \delta_{\bm{k}, -\bm{k}'}$,   $ \langle   \tilde{\eta}_{\bm{k}}\tilde{\xi}_{\bm{k}'}\rangle = 0$.
Combining these with Eq.~\myref{eq:u_k_solutions}, it immediately follows that
$\langle \tilde{u}_{\bm{k}}(t) \tilde{u}_{\bm{k}'}(t') \rangle = \langle \tilde{u}_{\bm{k}}(t) \tilde{u}_{-\bm{k}}(t') \rangle \delta_{\bm{k},-\bm{k}'}$.
Starting from Eq.~\myref{eq:fourier} and using the results from Eq.~\myref{eq:u_k_solutions}, we have
\begin{equation}
\begin{split}
&  \langle u(\bm{x},t) u(\bm{x}',t)  \rangle  = \frac{1}{L^{2d}}\sum_{\bm{k},\bm{k}'}   e^{i\bm{k \cdot x}} e^{i\bm{k}' \cdot \bm{x}'}
    \langle \tilde{u}_{\bm{k}}(t) \tilde{u}_{\bm{k}'}(t) \rangle
=  \frac{1}{L^{2d}}\sum_{\bm{k}} e^{i\bm{k} \cdot (\bm{x}-\bm{x}')}  \langle \tilde{u}_{\bm{k}}(t) \tilde{u}_{-\bm{k}}(t) \rangle  \\
& = \frac{1}{4L^{2d}}\sum_{\bm{k}} e^{i\bm{k} \cdot (\bm{x}-\bm{x}')}
\left\langle
\left( (\tilde{\eta}_{\bm{k}} - i\tilde{\xi}_{\bm{k}} ) e^{i\omega_{\bm{k}}t} +
(\tilde{\eta}_{\bm{k}} + i\tilde{\xi}_{\bm{k}} ) e^{-i\omega_{\bm{k}}t} \right)
\left( (\tilde{\eta}_{-\bm{k}} - i\tilde{\xi}_{-\bm{k}} ) e^{i\omega_{-\bm{k}}t} +
(\tilde{\eta}_{-\bm{k}} + i\tilde{\xi}_{-\bm{k}} ) e^{-i\omega_{-\bm{k}}t} \right)
\right\rangle \\
& = \frac{1}{4L^{d}}\sum_{\bm{k}} e^{i\bm{k} \cdot (\bm{x}-\bm{x}')}
\left( 2(D_u+D_{\vartheta}) +
(D_u-D_{\vartheta})( e^{2i\omega_{\bm{k}}t} + e^{-2i\omega_{\bm{k}}t} ) \right) \\
& \simeq \frac{1}{4}   \int\frac{d^dk}{(2\pi)^d} e^{i\bm{k} \cdot (\bm{x}-\bm{x}')}
\left( 2(D_u+D_{\vartheta}) +
(D_u-D_{\vartheta})( e^{2i\omega_{\bm{k}}t} + e^{-2i\omega_{\bm{k}}t} ) \right)
\;\;\;,
\end{split}
\label{u_sum}
\end{equation}
where we exploited that $\omega_{-\bm{k}}=\omega_{\bm{k}}=\alpha\bm{k}^2$ and in the large system-size limit we approximated the summation over each component of the wave number $k_j$ with an integral, $\sum_{k_j} (\ldots) \simeq \int_{-\pi/\Delta x}^{\pi/\Delta x}\frac{L}{2\pi}dk_j (\ldots)$ for a given spatial discretization $\Delta x$.
In the continuum limit, $\Delta x\to 0$, the upper/lower limits of the integral will approach $\pm \infty$. In this limit, the above two-point function can be written as
\begin{equation}
\begin{split}
& \langle u(\bm{x},t) u(\bm{x}',t)  \rangle  \simeq \\
& \frac{D_u+D_{\vartheta}}{2}\delta(\bm{x}-\bm{x}') + \frac{D_u-D_{\vartheta}}{4}  \int \frac{d^dk}{(2\pi)^d}
(e^{i\bm{k} \cdot (\bm{x}-\bm{x}')+2i\alpha \bm{k}^2 t} + e^{i\bm{k} \cdot (\bm{x}-\bm{x}') -2i\alpha \bm{k}^2 t})  \;\;\;.
\end{split}
\end{equation}
Completing the squares in the factorized exponentials, and employing the generalization of Gaussian integrals (including purely imaginary arguments) \cite{Townsend_2012},

\begin{equation}
\int_{-\infty}^{\infty} dy e^{\pm i\beta y^2} =\sqrt{\frac{\pi}{\mp i\beta}} = e^{\pm i\pi/4}\sqrt{\frac{\pi}{\beta}}
\end{equation}
(for $\beta>0$ and interpreting $\sqrt{i}=e^{i\pi/4}$), one has
\begin{equation}
\begin{split}
& \int_{-\infty}^{\infty} dk_j e^{i k_j (x_j-x_j') \pm 2i\alpha k_j^2 t} \\
& = \int_{-\infty}^{\infty} dk_j e^{\pm 2i\alpha t \left( k_j \pm\frac{x_j-x_j'}{4\alpha t}\right)^2 \mp i\frac{(x_j-x_j')^2}{8\alpha t} } =
e^{\mp i\frac{(x_j-x_j')^2}{8\alpha t} } \sqrt{\frac{\pi}{\mp 2i\alpha t}} = \sqrt{\frac{\pi}{2\alpha t}} e^{\mp i\frac{(x_j-x_j')^2}{8\alpha t} \pm i\frac{\pi}{4} }  \;\;\;.
\end{split}
\label{CGI}
\end{equation}
Hence, for the two-point correlation function in the large system size and continuum limit we have
\begin{equation}
\begin{split}
\langle u(\bm{x},t) u(\bm{x}',t)  \rangle  \simeq
\frac{D_{u} + D_{\vartheta}}{2} \delta(\bm{x}-\bm{x}') +
   \frac{D_{u} - D_{\vartheta}}{2} \frac{1}{(8 \pi\alpha t)^{d/2}} \cos( \frac{(\bm{x}-\bm{x}')^2}{8 \alpha t} - \frac{d\pi}{4}) \;\;\;.
   \label{u-u_spatial_corr}
   \end{split}
\end{equation}
For the other two correlation functions we similarly find
\begin{equation}
\begin{split}
& \langle \vartheta(\bm{x},t) \vartheta(\bm{x}',t)  \rangle  \simeq
\frac{D_{u} + D_{\vartheta}}{2} \delta(\bm{x}-\bm{x}')
   -\frac{D_{u} - D_{\vartheta}}{2} \frac{1}{(8 \pi\alpha t)^{d/2}} \cos( \frac{(\bm{x}-\bm{x}')^2}{8 \alpha t} - \frac{d\pi}{4}) \;\;\;, \\
& \langle u(\bm{x},t) \vartheta(\bm{x}',t)  \rangle  \simeq
 \frac{D_{u} - D_{\vartheta}}{2} \frac{1}{(8 \pi\alpha t)^{d/2}} \sin( \frac{(\bm{x}-\bm{x}')^2}{8 \alpha t} - \frac{d\pi}{4}) \;\;\;.
 \label{th-th_spatial_corr}
\end{split}
\end{equation}
One can recognize that the non-trivial parts (other than the delta-functions) in the above correlation functions are in the form of a real linear combination of the quantum propagator of a free particle \cite{Sakurai_2021,Townsend_2012}. One can also check that the initial conditions for the amplitude and phase variances and correlations are satisfied by using the limits
\begin{equation}
\delta(y)=\lim_{t \to 0} \frac{1}{\sqrt{\pm i 8\pi\alpha t }} e^{\pm i \frac{y^2}{8\alpha t} }  =
\lim_{t \to 0} \frac{1}{\sqrt{8\pi\alpha t }} e^{\pm i \frac{y^2}{8\alpha t} \mp i\pi/4}  \;\;\; .
\label{delta_function}
\end{equation}

With some additional effort, we can also approximate the spatial correlation functions for a finite discretization $\Delta x$. In this case, the summation in Eq.~\myref{u_sum} can still be approximated with an integral in the large-system size limit, but with an upper cutoff ($\pi/\Delta x$) in the magnitude of the momentum components $k_j$ in each dimension,
\begin{equation}
\begin{split}
& \int_{-\pi/\Delta x}^{\pi/\Delta x} dk_j e^{i k_j (x_j-x_j') \pm 2i\alpha k_j^2 t} \\
& = \sqrt{\frac{\pi}{\mp 8i\alpha t}} e^{\mp i\frac{(x_j-x_j')^2}{8\alpha t} }  \left[
\erf(\sqrt{\mp 2i\alpha t}(\frac{\pi}{\Delta x} \pm \frac{x_j-x_j'}{4\alpha t}) )  -
\erf(\sqrt{\mp 2i\alpha t}(-\frac{\pi}{\Delta x} \pm \frac{x_j-x_j'}{4\alpha t}) )
\right] \\
& = \sqrt{\frac{\pi}{ 8\alpha t}} e^{\mp i\frac{(x_j-x_j')^2}{8\alpha t} \pm i\frac{\pi}{4}}
\left[
\erf(\sqrt{\mp 2i\alpha t}(\frac{\pi}{\Delta x} \pm \frac{x_j-x_j'}{4\alpha t}) )  -
\erf(\sqrt{\mp 2i\alpha t}(-\frac{\pi}{\Delta x} \pm \frac{x_j-x_j'}{4\alpha t}) )
\right]  \;\;\;,
\end{split}
\label{CGI_discrete}
\end{equation}
where
\begin{equation}
\erf(w) = \frac{2}{\sqrt{\pi}}\int_{0}^{w} dz e^{-z^2}
\label{error_function}
\end{equation}
is the error function with a complex argument. Exploiting some of the properties of the error function, $\erf(-w)=-\erf(w)$ and $\overline{\erf(w)}=\erf(\overline{w})$, we then can write the two-point spatial correlation functions as
\begin{equation}
\begin{split}
& \langle u(\bm{x},t) u(\bm{x}',t)  \rangle  \simeq
\frac{D_{u} + D_{\vartheta}}{2} \frac{\delta_{\bm{x},\bm{x}'}}{(\Delta x)^d}
+  \frac{D_{u} - D_{\vartheta}}{2} \frac{1}{(8\pi\alpha t)^{d/2}}
   \Re \left( \prod_{j=1}^{d} F(x_j - x_j',t;\Delta x) \right)  \\
& \langle \vartheta(\bm{x},t) \vartheta(\bm{x}',t)  \rangle  \simeq
\frac{D_{u} + D_{\vartheta}}{2} \frac{\delta_{\bm{x},\bm{x}'}}{(\Delta x)^d}
- \frac{D_{u} - D_{\vartheta}}{2} \frac{1}{(8\pi\alpha t)^{d/2}}
   \Re \left( \prod_{j=1}^{d} F(x_j - x_j',t;\Delta x) \right)  \\
& \langle u(\bm{x},t) \vartheta(\bm{x}',t)  \rangle  \simeq
 \frac{D_{u} - D_{\vartheta}}{2} \frac{1}{(8\pi\alpha t)^{d/2}}
   \Im \left( \prod_{j=1}^{d} F(x_j - x_j',t;\Delta x) \right)
\end{split} \;\;\;,
\label{corr_spatial_discrete-app}
\end{equation}
where
\begin{equation}
F(y,t;\Delta x) \equiv
\frac{ e^{i\frac{y^2}{8\alpha t} -i\frac{\pi}{4} }  }{2}
\left[
\erf\left(\sqrt{2i\alpha t}(\frac{\pi}{\Delta x}+ \frac{y}{4\alpha t}) \right)
-
\erf\left(\sqrt{2i\alpha t}(-\frac{\pi}{\Delta x}+ \frac{y}{4\alpha t}) \right) \right]  \;\;\;.
\label{eq:F_y}
\end{equation}
Note that
\begin{equation}
\lim_{\Delta x\to 0} F(y,t;\Delta x) =  e^{ i\frac{y^2}{8\alpha t} - i\frac{\pi}{4} }   \;\;\;,
\end{equation}
hence Eqs.~\myref{u-u_spatial_corr} and \myref{th-th_spatial_corr} are recovered in the $\Delta x\to 0$ limit.
Further, employing the asymptotic expansion of the error function
\begin{equation}
\erf(w) \simeq 1- \frac{e^{-w^2}}{\sqrt{\pi}w}
\;\;\;,
\label{eq:error_func_asymp}
\end{equation}
we can obtain the lowest-order corrections to Eqs.~\myref{u-u_spatial_corr} and \myref{th-th_spatial_corr} for
$\frac{\alpha t}{(\Delta x)^2}\gg 1$. Here we focus on the single-site variance, hence $y = x_j - x_j' = 0$ for all $j$, yielding
\begin{equation}
\begin{split}
& F(0,t;\Delta x) =
\frac{ e^{-i\frac{\pi}{4} }  }{2}
\left[
\erf\left(\sqrt{2i\alpha t}(\frac{\pi}{\Delta x}) \right)
-
\erf\left(\sqrt{2i\alpha t}(-\frac{\pi}{\Delta x}) \right) \right]  \\
& = e^{-i\frac{\pi}{4}}
\erf\left(\sqrt{2i\alpha t}(\frac{\pi}{\Delta x}) \right)
\simeq
e^{-i\frac{\pi}{4}}
\left( 1 -  \frac{e^{-2i\alpha t(\frac{\pi}{\Delta x})^2}}
{\sqrt{2\pi i\alpha t}(\frac{\pi}{\Delta x})} \right)
\;\;\;.
\label{eq:F_0_asymp}
\end{split}
\end{equation}
Then for the amplitude variance we find
\begin{equation}
\begin{split}
& \sigma^2_u(t) =
\langle u^2(\bm{x},t)  \rangle  \simeq
\frac{D_{u} + D_{\vartheta}}{2} \frac{\delta_{\bm{x},\bm{x}'}}{(\Delta x)^d}
+  \frac{D_{u} - D_{\vartheta}}{2} \frac{1}{(8\pi\alpha t)^{d/2}}
   \Re \left(  F^d(0,t;\Delta x) \right)  \\
& \simeq \frac{D_{u} + D_{\vartheta}}{2} \frac{\delta_{\bm{x},\bm{x}'}}{(\Delta x)^d}
+  \frac{D_{u} - D_{\vartheta}}{2} \frac{1}{(8\pi\alpha t)^{d/2}} \left( \cos(\frac{d\pi}{4})
- \frac{d\cos(2\alpha t(\frac{\pi}{\Delta x})^2 +
\frac{(d+1)\pi}{4})}{\sqrt{2\pi \alpha t}(\frac{\pi}{\Delta x})}
\right) \;\;\; .
\label{eq:amplitude_variance_asymp}
\end{split}
\end{equation}
Similarly, for the phase variance we have
\begin{equation}
\begin{split}
& \sigma^2_{\vartheta}(t) =
\langle \vartheta^2(\bm{x},t)  \rangle  \simeq
\frac{D_{u} + D_{\vartheta}}{2} \frac{\delta_{\bm{x},\bm{x}'}}{(\Delta x)^d}
-  \frac{D_{u} - D_{\vartheta}}{2} \frac{1}{(8\pi\alpha t)^{d/2}}
   \Re \left(  F^d(0,t;\Delta x) \right)  \\
& \simeq \frac{D_{u} + D_{\vartheta}}{2} \frac{\delta_{\bm{x},\bm{x}'}}{(\Delta x)^d}
-  \frac{D_{u} - D_{\vartheta}}{2} \frac{1}{(8\pi\alpha t)^{d/2}} \left( \cos(\frac{d\pi}{4})
- \frac{d\cos(2\alpha t(\frac{\pi}{\Delta x})^2 +
\frac{(d+1)\pi}{4})}{\sqrt{2\pi \alpha t}(\frac{\pi}{\Delta x})}
\right) \;\;\; .
\label{eq:phase_variance_asymp}
\end{split}
\end{equation}

\section{Single-site temporal correlation functions (autocorrelations)}
\label{appendix_C}

As an example, we show the key steps in calculating the amplitude-amplitude single-site temporal correlation function. Exploiting again that in momentum space, $\langle \tilde{u}_{\bm{k}}(t) \tilde{u}_{\bm{k}'}(t') \rangle = \langle \tilde{u}_{\bm{k}}(t) \tilde{u}_{-\bm{k}}(t') \rangle \delta_{\bm{k},-\bm{k}'}$,
\begin{equation}
\begin{split}
&  \langle u(\bm{x},t) u(\bm{x},t')  \rangle  = \frac{1}{L^{2d}}\sum_{\bm{k},\bm{k}'}   e^{i\bm{k \cdot x}} e^{i\bm{k}' \cdot \bm{x}}
    \langle \tilde{u}_{\bm{k}}(t) \tilde{u}_{\bm{k}'}(t') \rangle
=  \frac{1}{L^{2d}}\sum_{\bm{k}} \langle \tilde{u}_{\bm{k}}(t) \tilde{u}_{-\bm{k}}(t') \rangle  \\
& = \frac{1}{4L^{2d}}\sum_{\bm{k}}
\left\langle
\left( (\tilde{\eta}_{\bm{k}} - i\tilde{\xi}_{\bm{k}} ) e^{i\omega_{\bm{k}}t} +
(\tilde{\eta}_{\bm{k}} + i\tilde{\xi}_{\bm{k}} ) e^{-i\omega_{\bm{k}}t} \right)
\left( (\tilde{\eta}_{-\bm{k}} - i\tilde{\xi}_{-\bm{k}} ) e^{i\omega_{-\bm{k}}t'} +
(\tilde{\eta}_{-\bm{k}} + i\tilde{\xi}_{-\bm{k}} ) e^{-i\omega_{-\bm{k}}t'} \right)
\right\rangle \\
& = \frac{1}{4L^{d}}\sum_{\bm{k}}
\left( (D_u+D_{\vartheta}) ( e^{i\omega_{\bm{k}}(t-t')} + e^{-i\omega_{\bm{k}}(t-t')} )
+
(D_u-D_{\vartheta})( e^{i\omega_{\bm{k}}(t+t')} + e^{-i\omega_{\bm{k}}(t+t')} ) \right) \\
& = \frac{1}{4L^{d}}\sum_{\bm{k}}
\left( (D_u+D_{\vartheta}) ( e^{i\alpha\bm{k}^2(t-t')} + e^{-i\alpha\bm{k}^2(t-t')} )
+
(D_u-D_{\vartheta})( e^{i\alpha\bm{k}^2(t+t')} + e^{-i\alpha\bm{k}^2(t+t')} ) \right) \\
& \simeq \frac{1}{4} \int \frac{d^dk}{(2\pi)^d}
\left( (D_u+D_{\vartheta}) ( e^{i\alpha\bm{k}^2(t-t')} + e^{-i\alpha\bm{k}^2(t-t')} )
+
(D_u-D_{\vartheta})( e^{i\alpha\bm{k}^2(t+t')} + e^{-i\alpha\bm{k}^2(t+t')} ) \right)  \;\;\;,
\end{split}
\label{u_temporal_sum}
\end{equation}
where we replaced the sums with integrals, $\sum_{k_j} (\ldots) \simeq \int_{-\pi/\Delta x}^{\pi/\Delta x}\frac{L}{2\pi}dk_j (\ldots)$ in the large system-size limit. These momentum-space integrals factorize for each dimension. First, in the spatial continuum limit, $\Delta x\to 0$, using again the generalization of Gaussian integrals \cite{Townsend_2012}
\begin{equation}
\int_{-\infty}^{\infty} dk_j e^{\pm i\alpha k_j^2 \tau}
= \sqrt{\frac{\pi}{\mp i\alpha \tau}}
= \sqrt{\frac{\pi}{\alpha \tau}} e^{\pm i\frac{\pi}{4} }  \;\;\;
\end{equation}
(for $\tau>0$), we find
		\begin{equation}
			\langle u(\bm{x}, t) u(\bm{x}, t')  \rangle
			\simeq \frac{D_{u} + D_{\vartheta}}{2}  \frac{\cos(\frac{d\pi}{4})}{(4\pi\alpha |t-t'|)^{d/2}}  +
            \frac{D_{u} - D_{\vartheta}}{2}  \frac{\cos(\frac{d\pi}{4})}{(4\pi\alpha (t+t'))^{d/2}}  \;\;\;,
            \label{u-u_temporal_corr}
\end{equation}
Similarly, for the other two temporal correlation functions one can find,
\begin{equation}
		\begin{split}
\langle \vartheta(\bm{x}, t) \vartheta(\bm{x}, t')  \rangle
			& \simeq  \frac{D_{u} + D_{\vartheta}}{2}  \frac{\cos(\frac{d\pi}{4})}{(4\pi\alpha |t-t'|)^{d/2}}  -
            \frac{D_{u} - D_{\vartheta}}{2}  \frac{\cos(\frac{d\pi}{4})}{(4\pi\alpha (t+t'))^{d/2}}  \\
\langle u(\bm{x}, t) \vartheta(\bm{x}, t')  \rangle
			& \simeq   \frac{D_{u} + D_{\vartheta}}{2}  \frac{\sin(\frac{d\pi}{4})}{(4\pi\alpha |t-t'|)^{d/2}} \mathrm{sgn}^d(t-t')
        - \frac{D_{u} - D_{\vartheta}}{2}  \frac{\sin(\frac{d\pi}{4})}{(4\pi\alpha (t+t'))^{d/2}}
		\end{split} \;\;\;,
  \label{th-th_temporal_corr}
		\end{equation}
where $\mathrm{sgn}(\tau)$ is the sign function. In the above (continuous-space) expressions, the $t\to t'$ limit appears to be highly singular. This can be reconciled with the interpretation of the delta-functions in Eq.~\myref{delta_function}, and Eqs.~\myref{u-u_temporal_corr} and \myref{th-th_temporal_corr} becoming equivalent with
Eqs.\myref{u-u_spatial_corr} and \myref{th-th_spatial_corr} for $\mathbf{x}=\mathbf{x}'$.

For a finite spatial discretization $\Delta x$, the upper and lower limits of the momentum-space integrals are finite, and similar to Eq.~\myref{CGI_discrete}, the integrals in Eq.~\myref{u_temporal_sum} then can be expressed in terms of the error function [Eq.~\myref{error_function}],
\begin{equation}
\begin{split}
& \int_{-\pi/\Delta x}^{\pi/\Delta x} dk_j e^{\pm i\alpha k_j^2 \tau} \\
& = \frac{1}{2} \sqrt{\frac{\pi}{\mp i\alpha \tau}}
\left[
\erf\left(\sqrt{\mp i\alpha \tau}(\frac{\pi}{\Delta x}) \right)  -
\erf\left(\sqrt{\mp i\alpha \tau}(-\frac{\pi}{\Delta x}) \right)
\right] \\
& = \sqrt{\frac{\pi}{\alpha \tau}} \frac{e^{\pm i\frac{\pi}{4}}}{2}
\left[
\erf\left(\sqrt{\mp i\alpha \tau}(\frac{\pi}{\Delta x}) \right)  -
\erf\left(\sqrt{\mp i\alpha \tau}(-\frac{\pi}{\Delta x}) \right)
\right] \\
& = \sqrt{\frac{\pi}{\alpha \tau}} e^{\pm i\frac{\pi}{4}}
\erf\left(\sqrt{\mp i\alpha \tau}(\frac{\pi}{\Delta x}) \right)
\;\;\;.
\end{split}
\end{equation}
We then finally obtain
\begin{equation}
\begin{split}
& \langle u(\bm{x},t) u(\bm{x},t')  \rangle  \simeq \\
& \frac{D_{u} + D_{\vartheta}}{2} \frac{1}{(4\pi\alpha |t-t'|)^{d/2}}   \Re \left( H^d(t-t';\Delta x) \right)
+
\frac{D_{u} - D_{\vartheta}}{2} \frac{1}{(4\pi\alpha (t+t'))^{d/2}}   \Re \left( H^d(t+t';\Delta x) \right)  \\
& \langle \vartheta(\bm{x},t) \vartheta(\bm{x},t')  \rangle  \simeq \\
& \frac{D_{u} + D_{\vartheta}}{2} \frac{1}{(4\pi\alpha |t-t'|)^{d/2}}   \Re \left( H^d(t-t';\Delta x) \right)
-
\frac{D_{u} - D_{\vartheta}}{2} \frac{1}{(4\pi\alpha (t+t'))^{d/2}}   \Re \left( H^d(t+t';\Delta x) \right)
\\
& \langle u(\bm{x},t) \vartheta(\bm{x},t')  \rangle  \simeq \\
& -\frac{D_{u} + D_{\vartheta}}{2} \frac{\mathrm{sgn}^{d}(t-t')}{(4\pi\alpha |t-t'|)^{d/2}}   \Im \left( H^d(t-t';\Delta x) \right)
+
\frac{D_{u} - D_{\vartheta}}{2} \frac{1}{(4\pi\alpha (t+t'))^{d/2}}   \Im \left( H^d(t+t';\Delta x) \right)
\end{split} \;\;\;,
\label{corr_temporal_discrete_app}
\end{equation}
where
\begin{equation}
H(\tau;\Delta x) \equiv
e^{-i\frac{\pi}{4} }
\erf\left(\sqrt{i\alpha\tau}(\frac{\pi}{\Delta x}) \right)  \;\;\;.
\end{equation}
Note that
\begin{equation}
\lim_{\Delta x\to 0} H(\tau;\Delta x) =  e^{ -i\frac{\pi}{4} }   \;\;\;,
\end{equation}
hence Eqs.~\myref{u-u_temporal_corr} and \myref{th-th_temporal_corr} are recovered in the $\Delta x\to 0$ limit.

Importantly, in the long-time limit, $t=t'+\tau$ with $\tau$ fixed and $t'\to \infty$, the second terms in the above expressions [Eqs.~\myref{corr_temporal_discrete_app}] vanish, hence the these correlation functions become time-homogeneous, i.e., will only depend on $\tau$,
$C_{u}(\tau,t')\equiv\langle u(\bm{x},t'+\tau) u(\bm{x},t')\rangle\simeq \hat{C}_{u}(\tau)$ and
$C_{\vartheta}(\tau,t')\equiv\langle \vartheta(\bm{x},t'+\tau) \vartheta(\bm{x},t')\rangle\simeq \hat{C}_{\vartheta}(\tau)$,
\begin{equation}
\begin{split}
\hat{C}_{u}(\tau) \simeq \hat{C}_{\vartheta}(\tau)
\simeq
\frac{D_{u} + D_{\vartheta}}{2} \frac{1}{(4\pi\alpha |\tau|)^{d/2}}   \Re \left( H^d(\tau;\Delta x) \right)
\;\;\;.
\end{split}
\end{equation}
Further, employing the asymptotic expansion of the error function [Eq.~\myref{eq:error_func_asymp}] for $\alpha \tau/(\Delta x)^2\gg 1$,
\begin{equation}
H(t;a) = e^{-i\frac{\pi}{4}}
\erf\left(\sqrt{i\alpha \tau}(\frac{\pi}{\Delta x}) \right)
\simeq
e^{-i\frac{\pi}{4}}
\left( 1 -  \frac{e^{-i\alpha \tau(\frac{\pi}{\Delta x})^2}}
{\sqrt{\pi i\alpha \tau}(\frac{\pi}{\Delta x})} \right)
\;\;\;,
\end{equation}
for the amplitude autocorrelations we find
\begin{equation}
\begin{split}
\hat{C}_{u}(\tau)
& \simeq
\frac{D_{u} + D_{\vartheta}}{2} \frac{1}{(4\pi\alpha \tau)^{d/2}} \left( \cos(\frac{d\pi}{4})
- \frac{d\cos(\alpha \tau(\frac{\pi}{\Delta x})^2 +
\frac{(d+1)\pi}{4})}{\sqrt{\pi \alpha \tau}(\frac{\pi}{\Delta x})}
\right) \;\;\; .
\end{split}
\label{autocorr_asymptotic}
\end{equation}

\section{Initial noise distribution with a lower length-scale cutoff}
\label{appendix_D}

The discretization length $\Delta x$ is an artifact of the spatial discretization scheme for solving Schrodinger's equation. At the same time, so far we have defined the initial noise to be uncorrelated beyond this length scale (or delta-correlated in the continuum limit). It can be important, however, to separate the discretization length $\Delta x$ (which can be taken to $0$) from a different physical lengthscale $a$ characterizing the initial distribution of the noise.

One simple way to implement this lower (physical) length-scale cutoff is to implement an upper cutoff in wave-number (or momentum) space, such that
$\tilde{\eta}_{\bm{k}} \equiv 0$ and  $\tilde{\xi}_{\bm{k}} \equiv 0$ for
$k>k_{\rm \max}\equiv\pi/a$,
while leaving the structure of the noise unchanged for $k\leq k_{\rm \max}$,
$\langle   \tilde{\eta}_{\bm{k}} \rangle = 0$,  $\langle   \tilde{\xi}_{\bm{k}} \rangle = 0$,
$ \langle   \tilde{\eta}_{\bm{k}}\tilde{\eta}_{\bm{k}'}\rangle = D_u L^d \delta_{\bm{k}, -\bm{k}'}$,  $ \langle   \tilde{\xi}_{\bm{k}}\tilde{\xi}_{\bm{k}'}\rangle = D_{\vartheta} L^d \delta_{\bm{k}, -\bm{k}'}$,   $ \langle   \tilde{\eta}_{\bm{k}}\tilde{\xi}_{\bm{k}'}\rangle = 0$.
Then  the initial spatial correlations (e.g., for the amplitude) become
\begin{equation}
\begin{split}
&  \langle u(\bm{x},0) u(\bm{x}',0)  \rangle  = \frac{1}{L^{2d}}\sum_{\bm{k},\bm{k}'}   e^{i\bm{k \cdot x}} e^{i\bm{k}' \cdot \bm{x}'}
    \langle \tilde{u}_{\bm{k}}(0) \tilde{u}_{\bm{k}'}(0) \rangle
= \frac{1}{L^{2d}}\sum_{\bm{k},\bm{k}'}   e^{i\bm{k \cdot x}} e^{i\bm{k}' \cdot \bm{x}'}
    \langle \tilde{\eta}_{\bm{k}} \tilde{\eta}_{\bm{k}'} \rangle   \\
& =  \frac{1}{L^{2d}}\sum_{|\bm{k}|\leq k_{\rm \max}} e^{i\bm{k} \cdot (\bm{x}-\bm{x}')}  D_{u}L^d  =
\frac{D_u}{L^{d}}\sum_{|\bm{k}|\leq k_{\rm \max}} e^{i\bm{k} \cdot (\bm{x}-\bm{x}')}
\simeq D_u \int_{|\bm{k}|\leq k_{\rm \max}} \frac{d^dk}{(2\pi)^d} e^{i\bm{k} \cdot (\bm{x}-\bm{x}')} \\
& = D_u \prod_{j=1}^{d} \int_{-k_{\rm \max}}^{k_{\rm \max}} \frac{dk_j}{2\pi} e^{ik_j (x_j-x'_j)} = D_u   \prod_{j=1}^{d} \frac{\sin(k_{\rm \max}(x_j-x'_j))}{\pi (x_j-x'_j)} \\
& = D_u   \prod_{j=1}^{d} \frac{\sin(\pi(x_j-x'_j)/a)}{\pi (x_j-x'_j)}
= D_u \prod_{j=1}^{d} \delta_{(a)}(x_j-x'_j) = D_u \delta_{(a)}(\bm{x}-\bm{x}')
\;\;\;,
\end{split}
\end{equation}
where we have defined
\begin{equation}
\delta_{(a)}(x)\equiv \frac{\sin(\pi x/a)}{\pi x}
\;\;\;,
\label{delta_limit_a}
\end{equation}
and its $d$-dimensional version
$\delta_{(a)}(\bm{x}-\bm{x}') = \prod_{j=1}^{d} \delta_{(a)}(x_j-x'_j)$.
Equation~\myref{delta_limit_a} is a well-known finite-length limit representation of the delta-function, $\lim_{a\to 0}\delta_{(a)}(x) = \delta(x)$. Further, for a fixed $a$,
\begin{equation}
\lim_{x\to 0} \delta_{(a)}(x) = \frac{1}{a}
\;\;\;.
\end{equation}

With the above ``renormalized" initial noise, one can repeat the steps carried out in Appendix~\ref{appendix_B} and \ref{appendix_C}, and realize that the only changes will be the replacement of
$\frac{\delta_{\bm{x},\bm{x}'}}{(\Delta x)^d}$ (or $\delta(\bm{x}-\bm{x}')$) with $\delta_{a}(\bm{x}-\bm{x}')$ and $\Delta x$ with $a$ in the expressions involving the error functions. We then immediately find for the equal-time two-point correlation function (e.g., for the amplitude)
\begin{equation}
\langle u(\bm{x},t) u(\bm{x}',t)  \rangle  \simeq
\frac{D_{u} + D_{\vartheta}}{2}  \delta_{a}(\bm{x}-\bm{x}')
+  \frac{D_{u} - D_{\vartheta}}{2} \frac{1}{(8\pi\alpha t)^{d/2}}
   \Re \left( \prod_{j=1}^{d} F(x_j-x'_j,t;a) \right)
\;\;\;,
\end{equation}
specifically, for the single-site variance,
\begin{equation}
\sigma^2_u(t) = \langle u^2(\bm{x},t)  \rangle  \simeq
\frac{D_{u} + D_{\vartheta}}{2}  \frac{1}{a^d}
+  \frac{D_{u} - D_{\vartheta}}{2} \frac{1}{(8\pi\alpha t)^{d/2}}
   \Re \left( F^d(0,t;a) \right)
\;\;\;.
\end{equation}
Similarly, for the single-site temporal correlation function (or autocorrelation function) one finds
\begin{equation}
\begin{split}
& \langle u(\bm{x},t) u(\bm{x},t')  \rangle  \simeq \\
& \frac{D_{u} + D_{\vartheta}}{2} \frac{1}{(4\pi\alpha |t-t'|)^{d/2}}   \Re \left( H^d(t-t';a) \right)
+
\frac{D_{u} - D_{\vartheta}}{2} \frac{1}{(4\pi\alpha (t+t'))^{d/2}}   \Re \left( H^d(t+t';a) \right)
\;\;\;.
\end{split}
\end{equation}
In the long-time limit, $t=t'+\tau$ with $\tau$ fixed and $t'\to \infty$, the autocorrelations become time-homogeneous, i.e., will only depend on $\tau$,
\begin{equation}
C_{u}(\tau,t')\equiv\langle u(\bm{x},t'+\tau) u(\bm{x},t')\rangle\simeq \hat{C}_{u}(\tau) \simeq
\frac{D_{u} + D_{\vartheta}}{2} \frac{1}{(4\pi\alpha |\tau|)^{d/2}}   \Re \left( H^d(\tau;a) \right)
\;\;\;.
\end{equation}
The asymptotic behavior of this autocorrelation function (analogous to Eq.~\myref{autocorr_asymptotic}) for $\alpha\tau/a^2\gg 1$ becomes
\begin{equation}
\begin{split}
\hat{C}_{u}(\tau)
& \simeq
\frac{D_{u} + D_{\vartheta}}{2} \frac{1}{(4\pi\alpha \tau)^{d/2}} \left( \cos(\frac{d\pi}{4})
- \frac{d\cos(\alpha \tau(\frac{\pi}{a})^2 +
\frac{(d+1)\pi}{4})}{\sqrt{\pi \alpha \tau}(\frac{\pi}{a})}
\right) \;\;\;.
\end{split}
\label{autocorr_asymptotic_cutoff_a}
\end{equation}

\bibliographystyle{apsrev4-2_CN}

\bibliography{manuscript}

\begin{thebibliography}{47}%
\makeatletter
\providecommand \@ifxundefined [1]{%
 \@ifx{#1\undefined}
}%
\providecommand \@ifnum [1]{%
 \ifnum #1\expandafter \@firstoftwo
 \else \expandafter \@secondoftwo
 \fi
}%
\providecommand \@ifx [1]{%
 \ifx #1\expandafter \@firstoftwo
 \else \expandafter \@secondoftwo
 \fi
}%
\providecommand \natexlab [1]{#1}%
\providecommand \enquote  [1]{``#1''}%
\providecommand \bibnamefont  [1]{#1}%
\providecommand \bibfnamefont [1]{#1}%
\providecommand \citenamefont [1]{#1}%
\providecommand \href@noop [0]{\@secondoftwo}%
\providecommand \href [0]{\begingroup \@sanitize@url \@href}%
\providecommand \@href[1]{\@@startlink{#1}\@@href}%
\providecommand \@@href[1]{\endgroup#1\@@endlink}%
\providecommand \@sanitize@url [0]{\catcode `\\12\catcode `\$12\catcode
  `\&12\catcode `\#12\catcode `\^12\catcode `\_12\catcode `\%12\relax}%
\providecommand \@@startlink[1]{}%
\providecommand \@@endlink[0]{}%
\providecommand \url  [0]{\begingroup\@sanitize@url \@url }%
\providecommand \@url [1]{\endgroup\@href {#1}{\urlprefix }}%
\providecommand \urlprefix  [0]{URL }%
\providecommand \Eprint [0]{\href }%
\providecommand \doibase [0]{https://doi.org/}%
\providecommand \selectlanguage [0]{\@gobble}%
\providecommand \bibinfo  [0]{\@secondoftwo}%
\providecommand \bibfield  [0]{\@secondoftwo}%
\providecommand \translation [1]{[#1]}%
\providecommand \BibitemOpen [0]{}%
\providecommand \bibitemStop [0]{}%
\providecommand \bibitemNoStop [0]{.\EOS\space}%
\providecommand \EOS [0]{\spacefactor3000\relax}%
\providecommand \BibitemShut  [1]{\csname bibitem#1\endcsname}%
\let\auto@bib@innerbib\@empty
\bibitem [{\citenamefont {Bray}\ \emph {et~al.}(2013)\citenamefont {Bray},
  \citenamefont {Majumdar},\ and\ \citenamefont {Schehr}}]{Bray2013}%
  \BibitemOpen
  \bibfield  {author} {\bibinfo {author} {\bibfnamefont {A.~J.}\ \bibnamefont
  {Bray}}, \bibinfo {author} {\bibfnamefont {S.~N.}\ \bibnamefont {Majumdar}},\
  and\ \bibinfo {author} {\bibfnamefont {G.}~\bibnamefont {Schehr}},\
  }\bibfield  {title} {\emph {\bibinfo {title} {Persistence and first-passage
  properties in nonequilibrium systems}},\ }\href
  {https://doi.org/10.1080/00018732.2013.803819} {\bibfield  {journal}
  {\bibinfo  {journal} {Advances in Physics}\ }\textbf {\bibinfo {volume}
  {62}},\ \bibinfo {pages} {225} (\bibinfo {year} {2013})}\BibitemShut
  {NoStop}%
\bibitem [{\citenamefont {Majumdar}\ \emph
  {et~al.}(1996{\natexlab{a}})\citenamefont {Majumdar}, \citenamefont {Sire},
  \citenamefont {Bray},\ and\ \citenamefont {Cornell}}]{Majumdar_PRL1996}%
  \BibitemOpen
  \bibfield  {author} {\bibinfo {author} {\bibfnamefont {S.~N.}\ \bibnamefont
  {Majumdar}}, \bibinfo {author} {\bibfnamefont {C.}~\bibnamefont {Sire}},
  \bibinfo {author} {\bibfnamefont {A.~J.}\ \bibnamefont {Bray}},\ and\
  \bibinfo {author} {\bibfnamefont {S.~J.}\ \bibnamefont {Cornell}},\
  }\bibfield  {title} {\emph {\bibinfo {title} {Nontrivial exponent for simple
  diffusion}},\ }\href {https://doi.org/10.1103/PhysRevLett.77.2867} {\bibfield
   {journal} {\bibinfo  {journal} {Phys. Rev. Lett.}\ }\textbf {\bibinfo
  {volume} {77}},\ \bibinfo {pages} {2867} (\bibinfo {year}
  {1996}{\natexlab{a}})}\BibitemShut {NoStop}%
\bibitem [{\citenamefont {Newman}\ and\ \citenamefont
  {Toroczkai}(1998)}]{newman98}%
  \BibitemOpen
  \bibfield  {author} {\bibinfo {author} {\bibfnamefont {T.~J.}\ \bibnamefont
  {Newman}}\ and\ \bibinfo {author} {\bibfnamefont {Z.}~\bibnamefont
  {Toroczkai}},\ }\bibfield  {title} {\emph {\bibinfo {title} {Diffusive
  persistence and the ``sign-time'' distribution}},\ }\href
  {https://doi.org/10.1103/PhysRevE.58.R2685} {\bibfield  {journal} {\bibinfo
  {journal} {Phys. Rev. E}\ }\textbf {\bibinfo {volume} {58}},\ \bibinfo
  {pages} {R2685} (\bibinfo {year} {1998})}\BibitemShut {NoStop}%
\bibitem [{\citenamefont {Ben-Naim}\ and\ \citenamefont
  {Krapivsky}(2014)}]{ben-naim2014}%
  \BibitemOpen
  \bibfield  {author} {\bibinfo {author} {\bibfnamefont {E.}~\bibnamefont
  {Ben-Naim}}\ and\ \bibinfo {author} {\bibfnamefont {P.~L.}\ \bibnamefont
  {Krapivsky}},\ }\bibfield  {title} {\emph {\bibinfo {title} {Slow kinetics of
  brownian maxima}},\ }\href {https://doi.org/10.1103/PhysRevLett.113.030604}
  {\bibfield  {journal} {\bibinfo  {journal} {Phys. Rev. Lett.}\ }\textbf
  {\bibinfo {volume} {113}},\ \bibinfo {pages} {030604} (\bibinfo {year}
  {2014})}\BibitemShut {NoStop}%
\bibitem [{\citenamefont {Derrida}\ \emph {et~al.}(1994)\citenamefont
  {Derrida}, \citenamefont {Bray},\ and\ \citenamefont
  {Godreche}}]{Derrida_JPA1994}%
  \BibitemOpen
  \bibfield  {author} {\bibinfo {author} {\bibfnamefont {B.}~\bibnamefont
  {Derrida}}, \bibinfo {author} {\bibfnamefont {A.~J.}\ \bibnamefont {Bray}},\
  and\ \bibinfo {author} {\bibfnamefont {C.}~\bibnamefont {Godreche}},\
  }\bibfield  {title} {\emph {\bibinfo {title} {Non-trivial exponents in the
  zero temperature dynamics of the 1d ising and potts models}},\ }\href
  {https://doi.org/10.1088/0305-4470/27/11/002} {\bibfield  {journal} {\bibinfo
   {journal} {Journal of Physics A: Mathematical and General}\ }\textbf
  {\bibinfo {volume} {27}},\ \bibinfo {pages} {L357} (\bibinfo {year}
  {1994})}\BibitemShut {NoStop}%
\bibitem [{\citenamefont {Derrida}\ \emph {et~al.}(1995)\citenamefont
  {Derrida}, \citenamefont {Hakim},\ and\ \citenamefont
  {Pasquier}}]{Derrida_PRL1995}%
  \BibitemOpen
  \bibfield  {author} {\bibinfo {author} {\bibfnamefont {B.}~\bibnamefont
  {Derrida}}, \bibinfo {author} {\bibfnamefont {V.}~\bibnamefont {Hakim}},\
  and\ \bibinfo {author} {\bibfnamefont {V.}~\bibnamefont {Pasquier}},\
  }\bibfield  {title} {\emph {\bibinfo {title} {Exact first-passage exponents
  of 1d domain growth: Relation to a reaction-diffusion model}},\ }\href
  {https://doi.org/10.1103/physrevlett.75.751} {\bibfield  {journal} {\bibinfo
  {journal} {Physical Review Letters}\ }\textbf {\bibinfo {volume} {75}},\
  \bibinfo {pages} {751–754} (\bibinfo {year} {1995})}\BibitemShut {NoStop}%
\bibitem [{\citenamefont {Howard}\ and\ \citenamefont
  {Godrèche}(1998)}]{Howard_JPA1998}%
  \BibitemOpen
  \bibfield  {author} {\bibinfo {author} {\bibfnamefont {M.}~\bibnamefont
  {Howard}}\ and\ \bibinfo {author} {\bibfnamefont {C.}~\bibnamefont
  {Godrèche}},\ }\bibfield  {title} {\emph {\bibinfo {title} {Persistence in
  the voter model: continuum reaction-diffusion approach}},\ }\href
  {https://doi.org/10.1088/0305-4470/31/11/001} {\bibfield  {journal} {\bibinfo
   {journal} {Journal of Physics A: Mathematical and General}\ }\textbf
  {\bibinfo {volume} {31}},\ \bibinfo {pages} {L209–L215} (\bibinfo {year}
  {1998})}\BibitemShut {NoStop}%
\bibitem [{\citenamefont {Silva}\ and\ \citenamefont
  {Dahmen}(2004)}]{Silva2004}%
  \BibitemOpen
  \bibfield  {author} {\bibinfo {author} {\bibfnamefont {R.~d.}\ \bibnamefont
  {Silva}}\ and\ \bibinfo {author} {\bibfnamefont {S.~R.}\ \bibnamefont
  {Dahmen}},\ }\bibfield  {title} {\emph {\bibinfo {title} {{Local persistence
  and blocking in the two-dimensional blume-capel model}}},\ }\href
  {https://doi.org/10.1590/S0103-97332004000700027} {\bibfield  {journal}
  {\bibinfo  {journal} {{Brazilian Journal of Physics}}\ }\textbf {\bibinfo
  {volume} {34}},\ \bibinfo {pages} {1469 } (\bibinfo {year}
  {2004})}\BibitemShut {NoStop}%
\bibitem [{\citenamefont {Majumdar}\ \emph
  {et~al.}(1996{\natexlab{b}})\citenamefont {Majumdar}, \citenamefont {Bray},
  \citenamefont {Cornell},\ and\ \citenamefont {Sire}}]{Majumdar_PRL1996b}%
  \BibitemOpen
  \bibfield  {author} {\bibinfo {author} {\bibfnamefont {S.~N.}\ \bibnamefont
  {Majumdar}}, \bibinfo {author} {\bibfnamefont {A.~J.}\ \bibnamefont {Bray}},
  \bibinfo {author} {\bibfnamefont {S.~J.}\ \bibnamefont {Cornell}},\ and\
  \bibinfo {author} {\bibfnamefont {C.}~\bibnamefont {Sire}},\ }\bibfield
  {title} {\emph {\bibinfo {title} {Global persistence exponent for
  nonequilibrium critical dynamics}},\ }\href
  {https://doi.org/10.1103/PhysRevLett.77.3704} {\bibfield  {journal} {\bibinfo
   {journal} {Phys. Rev. Lett.}\ }\textbf {\bibinfo {volume} {77}},\ \bibinfo
  {pages} {3704} (\bibinfo {year} {1996}{\natexlab{b}})}\BibitemShut {NoStop}%
\bibitem [{\citenamefont {Majumdar}\ and\ \citenamefont
  {Sire}(1996)}]{Majumdar_PRL1996c}%
  \BibitemOpen
  \bibfield  {author} {\bibinfo {author} {\bibfnamefont {S.~N.}\ \bibnamefont
  {Majumdar}}\ and\ \bibinfo {author} {\bibfnamefont {C.}~\bibnamefont
  {Sire}},\ }\bibfield  {title} {\emph {\bibinfo {title} {Survival probability
  of a gaussian non-markovian process: Application to the
  $\mathit{T}\phantom{\rule{0ex}{0ex}}=\phantom{\rule{0ex}{0ex}}0$ dynamics of
  the ising model}},\ }\href {https://doi.org/10.1103/PhysRevLett.77.1420}
  {\bibfield  {journal} {\bibinfo  {journal} {Phys. Rev. Lett.}\ }\textbf
  {\bibinfo {volume} {77}},\ \bibinfo {pages} {1420} (\bibinfo {year}
  {1996})}\BibitemShut {NoStop}%
\bibitem [{\citenamefont {Fuchs}\ \emph {et~al.}(2008)\citenamefont {Fuchs},
  \citenamefont {Schelter}, \citenamefont {Ginelli},\ and\ \citenamefont
  {Hinrichsen}}]{Fuchs_2008}%
  \BibitemOpen
  \bibfield  {author} {\bibinfo {author} {\bibfnamefont {J.}~\bibnamefont
  {Fuchs}}, \bibinfo {author} {\bibfnamefont {J.}~\bibnamefont {Schelter}},
  \bibinfo {author} {\bibfnamefont {F.}~\bibnamefont {Ginelli}},\ and\ \bibinfo
  {author} {\bibfnamefont {H.}~\bibnamefont {Hinrichsen}},\ }\bibfield  {title}
  {\emph {\bibinfo {title} {Local persistence in the directed percolation
  universality class}},\ }\href
  {https://doi.org/10.1088/1742-5468/2008/04/p04015} {\bibfield  {journal}
  {\bibinfo  {journal} {Journal of Statistical Mechanics: Theory and
  Experiment}\ }\textbf {\bibinfo {volume} {2008}},\ \bibinfo {pages} {P04015}
  (\bibinfo {year} {2008})}\BibitemShut {NoStop}%
\bibitem [{\citenamefont {Grassberger}(2009)}]{Grassberger_JSM2009}%
  \BibitemOpen
  \bibfield  {author} {\bibinfo {author} {\bibfnamefont {P.}~\bibnamefont
  {Grassberger}},\ }\bibfield  {title} {\emph {\bibinfo {title} {Local
  persistence in directed percolation}},\ }\href
  {https://doi.org/10.1088/1742-5468/2009/08/p08021} {\bibfield  {journal}
  {\bibinfo  {journal} {Journal of Statistical Mechanics: Theory and
  Experiment}\ }\textbf {\bibinfo {volume} {2009}},\ \bibinfo {pages} {P08021}
  (\bibinfo {year} {2009})}\BibitemShut {NoStop}%
\bibitem [{\citenamefont {Toroczkai}\ \emph {et~al.}(1999)\citenamefont
  {Toroczkai}, \citenamefont {Newman},\ and\ \citenamefont
  {Das~Sarma}}]{Toro_PRE1999}%
  \BibitemOpen
  \bibfield  {author} {\bibinfo {author} {\bibfnamefont {Z.}~\bibnamefont
  {Toroczkai}}, \bibinfo {author} {\bibfnamefont {T.~J.}\ \bibnamefont
  {Newman}},\ and\ \bibinfo {author} {\bibfnamefont {S.}~\bibnamefont
  {Das~Sarma}},\ }\bibfield  {title} {\emph {\bibinfo {title} {Sign-time
  distributions for interface growth}},\ }\href
  {https://doi.org/10.1103/PhysRevE.60.R1115} {\bibfield  {journal} {\bibinfo
  {journal} {Phys. Rev. E}\ }\textbf {\bibinfo {volume} {60}},\ \bibinfo
  {pages} {R1115} (\bibinfo {year} {1999})}\BibitemShut {NoStop}%
\bibitem [{\citenamefont {Krug}\ \emph {et~al.}(1997)\citenamefont {Krug},
  \citenamefont {Kallabis}, \citenamefont {Majumdar}, \citenamefont {Cornell},
  \citenamefont {Bray},\ and\ \citenamefont {Sire}}]{Krug_PRE1997}%
  \BibitemOpen
  \bibfield  {author} {\bibinfo {author} {\bibfnamefont {J.}~\bibnamefont
  {Krug}}, \bibinfo {author} {\bibfnamefont {H.}~\bibnamefont {Kallabis}},
  \bibinfo {author} {\bibfnamefont {S.~N.}\ \bibnamefont {Majumdar}}, \bibinfo
  {author} {\bibfnamefont {S.~J.}\ \bibnamefont {Cornell}}, \bibinfo {author}
  {\bibfnamefont {A.~J.}\ \bibnamefont {Bray}},\ and\ \bibinfo {author}
  {\bibfnamefont {C.}~\bibnamefont {Sire}},\ }\bibfield  {title} {\emph
  {\bibinfo {title} {Persistence exponents for fluctuating interfaces}},\
  }\href {https://doi.org/10.1103/physreve.56.2702} {\bibfield  {journal}
  {\bibinfo  {journal} {Physical Review E}\ }\textbf {\bibinfo {volume} {56}},\
  \bibinfo {pages} {2702–2712} (\bibinfo {year} {1997})}\BibitemShut
  {NoStop}%
\bibitem [{\citenamefont {Rice}(1944)}]{Rice_1944}%
  \BibitemOpen
  \bibfield  {author} {\bibinfo {author} {\bibfnamefont {S.~O.}\ \bibnamefont
  {Rice}},\ }\bibfield  {title} {\emph {\bibinfo {title} {Mathematical analysis
  of random noise}},\ }\href
  {https://doi.org/10.1002/j.1538-7305.1944.tb00874.x} {\bibfield  {journal}
  {\bibinfo  {journal} {The Bell System Technical Journal}\ }\textbf {\bibinfo
  {volume} {23}},\ \bibinfo {pages} {282} (\bibinfo {year} {1944})}\BibitemShut
  {NoStop}%
\bibitem [{\citenamefont {Rice}(1945)}]{Rice_1945}%
  \BibitemOpen
  \bibfield  {author} {\bibinfo {author} {\bibfnamefont {S.~O.}\ \bibnamefont
  {Rice}},\ }\bibfield  {title} {\emph {\bibinfo {title} {Mathematical analysis
  of random noise}},\ }\href
  {https://doi.org/10.1002/j.1538-7305.1945.tb00453.x} {\bibfield  {journal}
  {\bibinfo  {journal} {The Bell System Technical Journal}\ }\textbf {\bibinfo
  {volume} {24}},\ \bibinfo {pages} {46} (\bibinfo {year} {1945})}\BibitemShut
  {NoStop}%
\bibitem [{\citenamefont {Newell}\ and\ \citenamefont
  {Rosenblatt}(1962)}]{Newell_1962}%
  \BibitemOpen
  \bibfield  {author} {\bibinfo {author} {\bibfnamefont {G.~F.}\ \bibnamefont
  {Newell}}\ and\ \bibinfo {author} {\bibfnamefont {M.}~\bibnamefont
  {Rosenblatt}},\ }\bibfield  {title} {\emph {\bibinfo {title} {Zero crossing
  probabilities for gaussian stationary processes}},\ }\href
  {http://www.jstor.org/stable/2237990} {\bibfield  {journal} {\bibinfo
  {journal} {The Annals of Mathematical Statistics}\ }\textbf {\bibinfo
  {volume} {33}},\ \bibinfo {pages} {1306} (\bibinfo {year}
  {1962})}\BibitemShut {NoStop}%
\bibitem [{\citenamefont {Slepian}(1962)}]{Slepian_1962}%
  \BibitemOpen
  \bibfield  {author} {\bibinfo {author} {\bibfnamefont {D.}~\bibnamefont
  {Slepian}},\ }\bibfield  {title} {\emph {\bibinfo {title} {The one-sided
  barrier problem for gaussian noise}},\ }\href
  {https://doi.org/10.1002/j.1538-7305.1962.tb02419.x} {\bibfield  {journal}
  {\bibinfo  {journal} {The Bell System Technical Journal}\ }\textbf {\bibinfo
  {volume} {41}},\ \bibinfo {pages} {463} (\bibinfo {year} {1962})}\BibitemShut
  {NoStop}%
\bibitem [{\citenamefont {Derrida}\ \emph {et~al.}(1996)\citenamefont
  {Derrida}, \citenamefont {Hakim},\ and\ \citenamefont
  {Zeitak}}]{Derrida_PRL1996}%
  \BibitemOpen
  \bibfield  {author} {\bibinfo {author} {\bibfnamefont {B.}~\bibnamefont
  {Derrida}}, \bibinfo {author} {\bibfnamefont {V.}~\bibnamefont {Hakim}},\
  and\ \bibinfo {author} {\bibfnamefont {R.}~\bibnamefont {Zeitak}},\
  }\bibfield  {title} {\emph {\bibinfo {title} {Persistent spins in the linear
  diffusion approximation of phase ordering and zeros of stationary gaussian
  processes}},\ }\href {https://doi.org/10.1103/PhysRevLett.77.2871} {\bibfield
   {journal} {\bibinfo  {journal} {Phys. Rev. Lett.}\ }\textbf {\bibinfo
  {volume} {77}},\ \bibinfo {pages} {2871} (\bibinfo {year}
  {1996})}\BibitemShut {NoStop}%
\bibitem [{\citenamefont {Malik}\ \emph {et~al.}(2024)\citenamefont {Malik},
  \citenamefont {Varga}, \citenamefont {Moussawi}, \citenamefont {Hunt},
  \citenamefont {Szymanski}, \citenamefont {Toroczkai},\ and\ \citenamefont
  {Korniss}}]{Malik_PRE2024}%
  \BibitemOpen
  \bibfield  {author} {\bibinfo {author} {\bibfnamefont {O.}~\bibnamefont
  {Malik}}, \bibinfo {author} {\bibfnamefont {M.}~\bibnamefont {Varga}},
  \bibinfo {author} {\bibfnamefont {A.}~\bibnamefont {Moussawi}}, \bibinfo
  {author} {\bibfnamefont {D.}~\bibnamefont {Hunt}}, \bibinfo {author}
  {\bibfnamefont {B.}~\bibnamefont {Szymanski}}, \bibinfo {author}
  {\bibfnamefont {Z.}~\bibnamefont {Toroczkai}},\ and\ \bibinfo {author}
  {\bibfnamefont {G.}~\bibnamefont {Korniss}},\ }\bibfield  {title} {\emph
  {\bibinfo {title} {Diffusive persistence on disordered lattices and random
  networks}},\ }\href@noop {} {\bibfield  {journal} {\bibinfo  {journal} {Phys.
  Rev. E, in press}\ } (\bibinfo {year} {2024})},\ \bibinfo {note} {e-print
  arXiv:2207.05878}\BibitemShut {NoStop}%
\bibitem [{\citenamefont {Townsend}(2012)}]{Townsend_2012}%
  \BibitemOpen
  \bibfield  {author} {\bibinfo {author} {\bibfnamefont {J.}~\bibnamefont
  {Townsend}},\ }\href@noop {} {\emph {\bibinfo {title} {A Modern Approach to
  Quantum Mechanics}}},\ \bibinfo {edition} {2nd}\ ed.\ (\bibinfo  {publisher}
  {University Science Books},\ \bibinfo {year} {2012})\BibitemShut {NoStop}%
\bibitem [{\citenamefont {Poplavskyi}\ and\ \citenamefont
  {Schehr}(2018)}]{Schehr2018}%
  \BibitemOpen
  \bibfield  {author} {\bibinfo {author} {\bibfnamefont {M.}~\bibnamefont
  {Poplavskyi}}\ and\ \bibinfo {author} {\bibfnamefont {G.}~\bibnamefont
  {Schehr}},\ }\bibfield  {title} {\emph {\bibinfo {title} {Exact persistence
  exponent for the 2d-diffusion equation and related kac polynomials}},\ }\href
  {https://doi.org/10.1103/PhysRevLett.121.150601} {\bibfield  {journal}
  {\bibinfo  {journal} {Physical Review Letters}\ }\textbf {\bibinfo {volume}
  {121}},\ \bibinfo {pages} {150601} (\bibinfo {year} {2018})}\BibitemShut
  {NoStop}%
\bibitem [{\citenamefont {Dornic}(2018)}]{Dornic2018}%
  \BibitemOpen
  \bibfield  {author} {\bibinfo {author} {\bibfnamefont {I.}~\bibnamefont
  {Dornic}},\ }\href {https://doi.org/10.48550/ARXIV.1810.06957} {\bibinfo
  {title} {Universal {Painlevé} {VI} probability distribution in {Pfaffian}
  persistence and {Gaussian} first-passage problems with a sech-{Kernel}}}
  (\bibinfo {year} {2018}),\ \bibinfo {note} {e-print arXiv:1810.06957, 2018},\
  \Eprint {https://arxiv.org/abs/1810.06957} {arXiv:1810.06957} \BibitemShut
  {NoStop}%
\bibitem [{\citenamefont {Goswami}\ \emph {et~al.}(2010)\citenamefont
  {Goswami}, \citenamefont {Sen},\ and\ \citenamefont {Das}}]{Goswami_PRE2010}%
  \BibitemOpen
  \bibfield  {author} {\bibinfo {author} {\bibfnamefont {S.}~\bibnamefont
  {Goswami}}, \bibinfo {author} {\bibfnamefont {P.}~\bibnamefont {Sen}},\ and\
  \bibinfo {author} {\bibfnamefont {A.}~\bibnamefont {Das}},\ }\bibfield
  {title} {\emph {\bibinfo {title} {Quantum persistence: A random-walk
  scenario}},\ }\href {https://doi.org/10.1103/PhysRevE.81.021121} {\bibfield
  {journal} {\bibinfo  {journal} {Phys. Rev. E}\ }\textbf {\bibinfo {volume}
  {81}},\ \bibinfo {pages} {021121} (\bibinfo {year} {2010})}\BibitemShut
  {NoStop}%
\bibitem [{\citenamefont {Iitaka}(1994)}]{Iitaka_PRE1994}%
  \BibitemOpen
  \bibfield  {author} {\bibinfo {author} {\bibfnamefont {T.}~\bibnamefont
  {Iitaka}},\ }\bibfield  {title} {\emph {\bibinfo {title} {Solving the
  time-dependent schr\"odinger equation numerically}},\ }\href
  {https://doi.org/10.1103/PhysRevE.49.4684} {\bibfield  {journal} {\bibinfo
  {journal} {Physical Review E}\ }\textbf {\bibinfo {volume} {49}},\ \bibinfo
  {pages} {4684} (\bibinfo {year} {1994})}\BibitemShut {NoStop}%
\bibitem [{\citenamefont {Robertson}(2011)}]{Robertson_2011}%
  \BibitemOpen
  \bibfield  {author} {\bibinfo {author} {\bibfnamefont {D.~G.}\ \bibnamefont
  {Robertson}},\ }\href@noop {} {\bibinfo {title} {Solving the time-dependent
  schr\"odinger equation}} (\bibinfo {year} {2011})\BibitemShut {NoStop}%
\bibitem [{\citenamefont {Khan}\ \emph {et~al.}(2022)\citenamefont {Khan},
  \citenamefont {Ahsan}, \citenamefont {Bonyah}, \citenamefont {Jan},
  \citenamefont {Nisar}, \citenamefont {Abdel-Aty},\ and\ \citenamefont
  {Yahia}}]{Khan_2022}%
  \BibitemOpen
  \bibfield  {author} {\bibinfo {author} {\bibfnamefont {A.}~\bibnamefont
  {Khan}}, \bibinfo {author} {\bibfnamefont {M.}~\bibnamefont {Ahsan}},
  \bibinfo {author} {\bibfnamefont {E.}~\bibnamefont {Bonyah}}, \bibinfo
  {author} {\bibfnamefont {R.}~\bibnamefont {Jan}}, \bibinfo {author}
  {\bibfnamefont {M.}~\bibnamefont {Nisar}}, \bibinfo {author} {\bibfnamefont
  {A.-H.}\ \bibnamefont {Abdel-Aty}},\ and\ \bibinfo {author} {\bibfnamefont
  {I.~S.}\ \bibnamefont {Yahia}},\ }\bibfield  {title} {\emph {\bibinfo {title}
  {Numerical solution of schrödinger equation by crank–nicolson method}},\
  }\href {https://doi.org/10.1155/2022/6991067} {\bibfield  {journal} {\bibinfo
   {journal} {Mathematical Problems in Engineering}\ }\textbf {\bibinfo
  {volume} {2022}},\ \bibinfo {pages} {1–11} (\bibinfo {year}
  {2022})}\BibitemShut {NoStop}%
\bibitem [{\citenamefont {Sakurai}\ and\ \citenamefont
  {Napolitano}(2021)}]{Sakurai_2021}%
  \BibitemOpen
  \bibfield  {author} {\bibinfo {author} {\bibfnamefont {J.}~\bibnamefont
  {Sakurai}}\ and\ \bibinfo {author} {\bibfnamefont {J.}~\bibnamefont
  {Napolitano}},\ }\href@noop {} {\emph {\bibinfo {title} {Modern Quantum
  Mechanics}}},\ \bibinfo {edition} {3rd}\ ed.\ (\bibinfo  {publisher}
  {Cambridge University Press},\ \bibinfo {year} {2021})\BibitemShut {NoStop}%
\bibitem [{\citenamefont {Nore}\ \emph {et~al.}(1997)\citenamefont {Nore},
  \citenamefont {Abid},\ and\ \citenamefont {Brachet}}]{Nore_1997}%
  \BibitemOpen
  \bibfield  {author} {\bibinfo {author} {\bibfnamefont {C.}~\bibnamefont
  {Nore}}, \bibinfo {author} {\bibfnamefont {M.}~\bibnamefont {Abid}},\ and\
  \bibinfo {author} {\bibfnamefont {M.~E.}\ \bibnamefont {Brachet}},\
  }\bibfield  {title} {\emph {\bibinfo {title} {Decaying kolmogorov turbulence
  in a model of superflow}},\ }\href {https://doi.org/10.1063/1.869473}
  {\bibfield  {journal} {\bibinfo  {journal} {Physics of Fluids}\ }\textbf
  {\bibinfo {volume} {9}},\ \bibinfo {pages} {2644} (\bibinfo {year}
  {1997})}\BibitemShut {NoStop}%
\bibitem [{\citenamefont {Chiueh}\ \emph {et~al.}(2011)\citenamefont {Chiueh},
  \citenamefont {Woo}, \citenamefont {Jian},\ and\ \citenamefont
  {Schive}}]{Chiueh_2011}%
  \BibitemOpen
  \bibfield  {author} {\bibinfo {author} {\bibfnamefont {T.}~\bibnamefont
  {Chiueh}}, \bibinfo {author} {\bibfnamefont {T.-P.}\ \bibnamefont {Woo}},
  \bibinfo {author} {\bibfnamefont {H.-Y.}\ \bibnamefont {Jian}},\ and\
  \bibinfo {author} {\bibfnamefont {H.-Y.}\ \bibnamefont {Schive}},\ }\bibfield
   {title} {\emph {\bibinfo {title} {Vortex turbulence in linear schrödinger
  wave mechanics}},\ }\href {https://doi.org/10.1088/0953-4075/44/11/115101}
  {\bibfield  {journal} {\bibinfo  {journal} {Journal of Physics B: Atomic,
  Molecular and Optical Physics}\ }\textbf {\bibinfo {volume} {44}},\ \bibinfo
  {pages} {115101} (\bibinfo {year} {2011})}\BibitemShut {NoStop}%
\bibitem [{\citenamefont {Rogel-Salazar}(2013)}]{Salazar_EJP2013}%
  \BibitemOpen
  \bibfield  {author} {\bibinfo {author} {\bibfnamefont {J.}~\bibnamefont
  {Rogel-Salazar}},\ }\bibfield  {title} {\emph {\bibinfo {title} {The
  gross-pitaevskii equation and bose-einstein condensates}},\ }\href
  {https://doi.org/10.1088/0143-0807/34/2/247} {\bibfield  {journal} {\bibinfo
  {journal} {European Journal of Physics}\ }\textbf {\bibinfo {volume} {34}},\
  \bibinfo {pages} {247} (\bibinfo {year} {2013})}\BibitemShut {NoStop}%
\bibitem [{\citenamefont {Hellweg}\ \emph {et~al.}(2001)\citenamefont
  {Hellweg}, \citenamefont {Dettmer}, \citenamefont {Ryytty}, \citenamefont
  {Arlt}, \citenamefont {Ertmer}, \citenamefont {Sengstock}, \citenamefont
  {Petrov}, \citenamefont {Shlyapnikov}, \citenamefont {Kreutzmann},
  \citenamefont {Santos},\ and\ \citenamefont {Lewenstein}}]{Hellweg_APB2001}%
  \BibitemOpen
  \bibfield  {author} {\bibinfo {author} {\bibfnamefont {D.}~\bibnamefont
  {Hellweg}}, \bibinfo {author} {\bibfnamefont {S.}~\bibnamefont {Dettmer}},
  \bibinfo {author} {\bibfnamefont {P.}~\bibnamefont {Ryytty}}, \bibinfo
  {author} {\bibfnamefont {J.}~\bibnamefont {Arlt}}, \bibinfo {author}
  {\bibfnamefont {W.}~\bibnamefont {Ertmer}}, \bibinfo {author} {\bibfnamefont
  {K.}~\bibnamefont {Sengstock}}, \bibinfo {author} {\bibfnamefont
  {D.}~\bibnamefont {Petrov}}, \bibinfo {author} {\bibfnamefont
  {G.}~\bibnamefont {Shlyapnikov}}, \bibinfo {author} {\bibfnamefont
  {H.}~\bibnamefont {Kreutzmann}}, \bibinfo {author} {\bibfnamefont
  {L.}~\bibnamefont {Santos}},\ and\ \bibinfo {author} {\bibfnamefont
  {M.}~\bibnamefont {Lewenstein}},\ }\bibfield  {title} {\emph {\bibinfo
  {title} {Phase fluctuations in bose-einstein condensates}},\ }\href
  {https://doi.org/10.1007/s003400100747} {\bibfield  {journal} {\bibinfo
  {journal} {Applied Physics B}\ }\textbf {\bibinfo {volume} {73}},\ \bibinfo
  {pages} {781} (\bibinfo {year} {2001})}\BibitemShut {NoStop}%
\bibitem [{\citenamefont {Feldheim}\ and\ \citenamefont
  {Feldheim}(2014)}]{Feldheim_2014}%
  \BibitemOpen
  \bibfield  {author} {\bibinfo {author} {\bibfnamefont {N.~D.}\ \bibnamefont
  {Feldheim}}\ and\ \bibinfo {author} {\bibfnamefont {O.~N.}\ \bibnamefont
  {Feldheim}},\ }\bibfield  {title} {\emph {\bibinfo {title} {Long gaps between
  sign-changes of gaussian stationary processes}},\ }\bibfield  {journal}
  {\bibinfo  {journal} {International Mathematics Research Notices}\ }\href
  {https://doi.org/10.1093/imrn/rnu020} {10.1093/imrn/rnu020} (\bibinfo {year}
  {2014})\BibitemShut {NoStop}%
\bibitem [{\citenamefont {Dembo}\ and\ \citenamefont
  {Mukherjee}(2016)}]{Dembo_2016}%
  \BibitemOpen
  \bibfield  {author} {\bibinfo {author} {\bibfnamefont {A.}~\bibnamefont
  {Dembo}}\ and\ \bibinfo {author} {\bibfnamefont {S.}~\bibnamefont
  {Mukherjee}},\ }\bibfield  {title} {\emph {\bibinfo {title} {Persistence of
  gaussian processes: non-summable correlations}},\ }\href
  {https://doi.org/10.1007/s00440-016-0746-9} {\bibfield  {journal} {\bibinfo
  {journal} {Probability Theory and Related Fields}\ }\textbf {\bibinfo
  {volume} {169}},\ \bibinfo {pages} {1007} (\bibinfo {year}
  {2016})}\BibitemShut {NoStop}%
\bibitem [{\citenamefont {Feldheim}\ \emph {et~al.}(2021)\citenamefont
  {Feldheim}, \citenamefont {Feldheim},\ and\ \citenamefont
  {Mukherjee}}]{Feldheim_arxiv2021}%
  \BibitemOpen
  \bibfield  {author} {\bibinfo {author} {\bibfnamefont {N.}~\bibnamefont
  {Feldheim}}, \bibinfo {author} {\bibfnamefont {O.}~\bibnamefont {Feldheim}},\
  and\ \bibinfo {author} {\bibfnamefont {S.}~\bibnamefont {Mukherjee}},\
  }\href@noop {} {\bibinfo {title} {Persistence and ball exponents for gaussian
  stationary processes}} (\bibinfo {year} {2021}),\ \Eprint
  {https://arxiv.org/abs/2112.04820} {arXiv:2112.04820 [math.PR]} \BibitemShut
  {NoStop}%
\bibitem [{\citenamefont {Newman}\ and\ \citenamefont
  {Stein}(1999)}]{Newman1999}%
  \BibitemOpen
  \bibfield  {author} {\bibinfo {author} {\bibfnamefont {C.~M.}\ \bibnamefont
  {Newman}}\ and\ \bibinfo {author} {\bibfnamefont {D.~L.}\ \bibnamefont
  {Stein}},\ }\bibfield  {title} {\emph {\bibinfo {title} {Blocking and
  persistence in the zero-temperature dynamics of homogeneous and disordered
  ising models}},\ }\href {https://doi.org/10.1103/PhysRevLett.82.3944}
  {\bibfield  {journal} {\bibinfo  {journal} {Phys. Rev. Lett.}\ }\textbf
  {\bibinfo {volume} {82}},\ \bibinfo {pages} {3944} (\bibinfo {year}
  {1999})}\BibitemShut {NoStop}%
\bibitem [{\citenamefont {Anderson}(1958)}]{Anderson_PR1958}%
  \BibitemOpen
  \bibfield  {author} {\bibinfo {author} {\bibfnamefont {P.~W.}\ \bibnamefont
  {Anderson}},\ }\bibfield  {title} {\emph {\bibinfo {title} {Absence of
  {{Diffusion}} in {{Certain Random Lattices}}}},\ }\href
  {https://doi.org/10.1103/PhysRev.109.1492} {\bibfield  {journal} {\bibinfo
  {journal} {Physical Review}\ }\textbf {\bibinfo {volume} {109}},\ \bibinfo
  {pages} {1492} (\bibinfo {year} {1958})}\BibitemShut {NoStop}%
\bibitem [{\citenamefont {Domany}\ \emph {et~al.}(1983)\citenamefont {Domany},
  \citenamefont {Alexander}, \citenamefont {Bensimon},\ and\ \citenamefont
  {Kadanoff}}]{Domany_PRB1983}%
  \BibitemOpen
  \bibfield  {author} {\bibinfo {author} {\bibfnamefont {E.}~\bibnamefont
  {Domany}}, \bibinfo {author} {\bibfnamefont {S.}~\bibnamefont {Alexander}},
  \bibinfo {author} {\bibfnamefont {D.}~\bibnamefont {Bensimon}},\ and\
  \bibinfo {author} {\bibfnamefont {L.~P.}\ \bibnamefont {Kadanoff}},\
  }\bibfield  {title} {\emph {\bibinfo {title} {Solutions to the schr\"odinger
  equation on some fractal lattices}},\ }\href
  {https://doi.org/10.1103/PhysRevB.28.3110} {\bibfield  {journal} {\bibinfo
  {journal} {Phys. Rev. B}\ }\textbf {\bibinfo {volume} {28}},\ \bibinfo
  {pages} {3110} (\bibinfo {year} {1983})}\BibitemShut {NoStop}%
\bibitem [{\citenamefont {Soukoulis}\ \emph {et~al.}(1992)\citenamefont
  {Soukoulis}, \citenamefont {Li},\ and\ \citenamefont
  {Grest}}]{Soukoulis_PRB1992}%
  \BibitemOpen
  \bibfield  {author} {\bibinfo {author} {\bibfnamefont {C.~M.}\ \bibnamefont
  {Soukoulis}}, \bibinfo {author} {\bibfnamefont {Q.}~\bibnamefont {Li}},\ and\
  \bibinfo {author} {\bibfnamefont {G.~S.}\ \bibnamefont {Grest}},\ }\bibfield
  {title} {\emph {\bibinfo {title} {Quantum percolation in three-dimensional
  systems}},\ }\href {https://doi.org/10.1103/PhysRevB.45.7724} {\bibfield
  {journal} {\bibinfo  {journal} {Physical Review B}\ }\textbf {\bibinfo
  {volume} {45}},\ \bibinfo {pages} {7724} (\bibinfo {year}
  {1992})}\BibitemShut {NoStop}%
\bibitem [{\citenamefont {Wang}\ \emph {et~al.}(2013)\citenamefont {Wang},
  \citenamefont {Zhou}, \citenamefont {Zhang}, \citenamefont {Garoni},\ and\
  \citenamefont {Deng}}]{Wang_PRE2013}%
  \BibitemOpen
  \bibfield  {author} {\bibinfo {author} {\bibfnamefont {J.}~\bibnamefont
  {Wang}}, \bibinfo {author} {\bibfnamefont {Z.}~\bibnamefont {Zhou}}, \bibinfo
  {author} {\bibfnamefont {W.}~\bibnamefont {Zhang}}, \bibinfo {author}
  {\bibfnamefont {T.~M.}\ \bibnamefont {Garoni}},\ and\ \bibinfo {author}
  {\bibfnamefont {Y.}~\bibnamefont {Deng}},\ }\bibfield  {title} {\emph
  {\bibinfo {title} {Bond and site percolation in three dimensions}},\ }\href
  {https://doi.org/10.1103/PhysRevE.87.052107} {\bibfield  {journal} {\bibinfo
  {journal} {Phys. Rev. E}\ }\textbf {\bibinfo {volume} {87}},\ \bibinfo
  {pages} {052107} (\bibinfo {year} {2013})}\BibitemShut {NoStop}%
\bibitem [{\citenamefont {Zhu}\ and\ \citenamefont
  {Xiong}(2000)}]{Zhu_PRB2000}%
  \BibitemOpen
  \bibfield  {author} {\bibinfo {author} {\bibfnamefont {C.-P.}\ \bibnamefont
  {Zhu}}\ and\ \bibinfo {author} {\bibfnamefont {S.-J.}\ \bibnamefont
  {Xiong}},\ }\bibfield  {title} {\emph {\bibinfo {title}
  {Localization-delocalization transition of electron states in a disordered
  quantum small-world network}},\ }\href
  {https://doi.org/10.1103/PhysRevB.62.14780} {\bibfield  {journal} {\bibinfo
  {journal} {Phys. Rev. B}\ }\textbf {\bibinfo {volume} {62}},\ \bibinfo
  {pages} {14780} (\bibinfo {year} {2000})}\BibitemShut {NoStop}%
\bibitem [{\citenamefont {Kim}\ \emph {et~al.}(2003)\citenamefont {Kim},
  \citenamefont {Hong},\ and\ \citenamefont {Choi}}]{Kim_PRB2003}%
  \BibitemOpen
  \bibfield  {author} {\bibinfo {author} {\bibfnamefont {B.~J.}\ \bibnamefont
  {Kim}}, \bibinfo {author} {\bibfnamefont {H.}~\bibnamefont {Hong}},\ and\
  \bibinfo {author} {\bibfnamefont {M.~Y.}\ \bibnamefont {Choi}},\ }\bibfield
  {title} {\emph {\bibinfo {title} {Quantum and classical diffusion on
  small-world networks}},\ }\href {https://doi.org/10.1103/PhysRevB.68.014304}
  {\bibfield  {journal} {\bibinfo  {journal} {Physical Review B}\ }\textbf
  {\bibinfo {volume} {68}},\ \bibinfo {pages} {014304} (\bibinfo {year}
  {2003})}\BibitemShut {NoStop}%
\bibitem [{\citenamefont {M\"ulken}\ \emph {et~al.}(2007)\citenamefont
  {M\"ulken}, \citenamefont {Pernice},\ and\ \citenamefont
  {Blumen}}]{Mulken_PRE2007}%
  \BibitemOpen
  \bibfield  {author} {\bibinfo {author} {\bibfnamefont {O.}~\bibnamefont
  {M\"ulken}}, \bibinfo {author} {\bibfnamefont {V.}~\bibnamefont {Pernice}},\
  and\ \bibinfo {author} {\bibfnamefont {A.}~\bibnamefont {Blumen}},\
  }\bibfield  {title} {\emph {\bibinfo {title} {Quantum transport on
  small-world networks: A continuous-time quantum walk approach}},\ }\href
  {https://doi.org/10.1103/PhysRevE.76.051125} {\bibfield  {journal} {\bibinfo
  {journal} {Phys. Rev. E}\ }\textbf {\bibinfo {volume} {76}},\ \bibinfo
  {pages} {051125} (\bibinfo {year} {2007})}\BibitemShut {NoStop}%
\bibitem [{\citenamefont {Novotny}\ \emph {et~al.}(2014)\citenamefont
  {Novotny}, \citenamefont {Solomon},\ and\ \citenamefont
  {Inkoom}}]{Novotny_Procedia2014}%
  \BibitemOpen
  \bibfield  {author} {\bibinfo {author} {\bibfnamefont {M.}~\bibnamefont
  {Novotny}}, \bibinfo {author} {\bibfnamefont {L.}~\bibnamefont {Solomon}},\
  and\ \bibinfo {author} {\bibfnamefont {G.}~\bibnamefont {Inkoom}},\
  }\bibfield  {title} {\emph {\bibinfo {title} {Quantum transport through a
  fully connected network with disorder}},\ }\href
  {https://doi.org/https://doi.org/10.1016/j.phpro.2014.06.029} {\bibfield
  {journal} {\bibinfo  {journal} {Physics Procedia}\ }\textbf {\bibinfo
  {volume} {53}},\ \bibinfo {pages} {71} (\bibinfo {year} {2014})},\ \bibinfo
  {note} {26th Annual CSP Workshop on “Recent Developments in Computer
  Simulation Studies in Condensed Matter Physics”, CSP 2013}\BibitemShut
  {NoStop}%
\bibitem [{\citenamefont {Katzgraber}\ and\ \citenamefont
  {Novotny}(2018)}]{Novotny_PRA2018}%
  \BibitemOpen
  \bibfield  {author} {\bibinfo {author} {\bibfnamefont {H.~G.}\ \bibnamefont
  {Katzgraber}}\ and\ \bibinfo {author} {\bibfnamefont {M.}~\bibnamefont
  {Novotny}},\ }\bibfield  {title} {\emph {\bibinfo {title} {How small-world
  interactions can lead to improved quantum annealer designs}},\ }\href
  {https://doi.org/10.1103/PhysRevApplied.10.054004} {\bibfield  {journal}
  {\bibinfo  {journal} {Phys. Rev. Appl.}\ }\textbf {\bibinfo {volume} {10}},\
  \bibinfo {pages} {054004} (\bibinfo {year} {2018})}\BibitemShut {NoStop}%
\bibitem [{\citenamefont {Moro}\ \emph {et~al.}(2018)\citenamefont {Moro},
  \citenamefont {Dall'Osto},\ and\ \citenamefont {Fresch}}]{Moro_ChemPhys2018}%
  \BibitemOpen
  \bibfield  {author} {\bibinfo {author} {\bibfnamefont {G.~J.}\ \bibnamefont
  {Moro}}, \bibinfo {author} {\bibfnamefont {G.}~\bibnamefont {Dall'Osto}},\
  and\ \bibinfo {author} {\bibfnamefont {B.}~\bibnamefont {Fresch}},\
  }\bibfield  {title} {\emph {\bibinfo {title} {Signatures of anderson
  localization and delocalized random quantum states}},\ }\href
  {https://doi.org/https://doi.org/10.1016/j.chemphys.2018.03.006} {\bibfield
  {journal} {\bibinfo  {journal} {Chemical Physics}\ }\textbf {\bibinfo
  {volume} {514}},\ \bibinfo {pages} {141} (\bibinfo {year} {2018})},\ \bibinfo
  {note} {energy and Entropy of Change: From Elementary Processes to
  Biology}\BibitemShut {NoStop}%
\bibitem [{\citenamefont {Ma}\ \emph {et~al.}(2024)\citenamefont {Ma},
  \citenamefont {Malik},\ and\ \citenamefont {Korniss}}]{Korniss_2024}%
  \BibitemOpen
  \bibfield  {author} {\bibinfo {author} {\bibfnamefont {C.}~\bibnamefont
  {Ma}}, \bibinfo {author} {\bibfnamefont {O.}~\bibnamefont {Malik}},\ and\
  \bibinfo {author} {\bibfnamefont {G.}~\bibnamefont {Korniss}}} (\bibinfo
  {year} {2024}),\ \bibinfo {note} {in preparation}\BibitemShut {NoStop}%
\end{thebibliography}%

\end{document}